\documentclass[prd,aps,showpacs,twocolumn,superscriptaddress,floatfix,nofootinbib]{revtex4-1}
\usepackage[utf8]{inputenc}
\usepackage{bm}
\usepackage{mathtools}
\usepackage[toc,page]{appendix}
\usepackage{dsfont}
\usepackage[colorlinks=true,linkcolor=blue,citecolor=blue, urlcolor=blue]{hyperref} 
\usepackage[a4paper,top=2.4cm,bottom=2.4cm,right=1.0cm,bindingoffset=-1.2cm]{geometry}
\usepackage{natbib}
\usepackage{tikz}
\usepackage{graphicx}
\usepackage{subfigure}
\usepackage{scalerel,amssymb}
\begin{document}
\raggedbottom
\title{3-D structure of the Glasma initial state -- Breaking boost-invariance by collisions of extended shock waves in classical Yang-Mills theory}

\author{S.~Schlichting}
\affiliation{Fakultät für Physik, Universität Bielefeld, D-33615 Bielefeld, Germany}
\author{P.~Singh}
\email{pragya@physik.uni-bielefeld.de}
\affiliation{Fakultät für Physik, Universität Bielefeld, D-33615 Bielefeld, Germany}
\date{\today}

\begin{abstract}
We simulate the 3+1 D classical Yang-Mills dynamics of the collisions of longitudinally extended nuclei, described by eikonal color charges in the Color Glass Condensate framework. By varying the longitudinal thickness of the colliding nuclei, we discuss the violations of boost invariance and explore how the boost invariant high-energy limit is approached. Subsequently, we develop a more realistic model of the three dimensional color charge distributions that connects longitudinal and transverse fluctuations to the $x$ and $\mathbf{k_\bot}$ dependence of transverse momentum dependent parton distributions, and explore the resulting rapidity profiles and their longitudinal fluctuations. 
\end{abstract}

\pacs{}
\maketitle
\section{Introduction}
High-energy heavy-ion collisions at the Large Hadron Collider (LHC) and the Relativistic Heavy Ion Collider (RHIC) produce a deconfined Quark Gluon Plasma (QGP), whose space-time dynamics is well described by relativistic viscous hydrodynamics \cite{Gale:2013da,Heinz:2013th}. Over the course of the last few years sophisticated multi-stage models, including the early time pre-equilibrium dynamics, the intermediate hydrodynamic QGP phase, as well as the late stage hadronic re-scattering have been developed to describe the space-time dynamics of heavy-ion collisions. Nevertheless, a comprehensive understanding hinges to a considerable extent on a proper theoretical understanding of the initial conditions, characterizing the non-equilibrium QGP created immediately after the collision of heavy nuclei.

The Color Glass Condensate (CGC) effective theory of high-energy QCD \cite{Gelis:2010nm,Gelis:2012ri} provides a framework to study the earliest stages of the collision which encompasses the description of dense colliding nuclei prior to the collision and their initial energy deposition \cite{Kovner:1995ja,Krasnitz:1998ns}. Within this framework, the  high-energy boost-invariant limit has been studied in great detail \cite{Krasnitz:1998ns,Krasnitz:1997zj,Lappi:2003bi,Lappi:2006fp} and lead to the successful development of microscopic initial state models, such as IP-Glasma \cite{Schenke:2012wb,Schenke:2012hg} and MC-KLN \cite{Drescher:2006ca,Drescher:2007ax}, which describe the two dimensional transverse energy density profiles. Beyond numerical studies, a lot of (semi) analytical CGC calculations have been performed to
characterize the correlation functions of the initial energy-momentum tensor \cite{Lappi:2017skr,Albacete:2018bbv,Giacalone:2019kgg} and understand the properties of the boost-invariant Glasma at early times \cite{Fries:2005yc,Fukushima:2007yk,Fujii:2008km,Chen:2015wia}.
    
Based on new experimental measurements at lower energies and interesting results of experimental measurements of longitudinal fluctuations at high energies, there has recently been a renewed interest in understanding the three dimensional structure of the initial state, that has lead to the development of various three dimensional initial state models \cite{Bozek:2015bna,Bozek:2010vz,Pang:2016akb,Shen:2020jwv}.
However, so far only a limited number of first principles insights have been obtained beyond the  high-energy boost invariant limit, with existing analytic studies focusing on the fragmentation region \cite{Kolbe:2020hem,Lushozi:2019duv,Kajantie:2019nse,Kajantie:2019hft,McLerran:2018avb} and sub-eikonal corrections to particle production in momentum space \cite{Altinoluk:2020wpf,Altinoluk:2015qoa}. Beyond that, there is a number of proposals that aim to generalize the boost-invariant IP-Glasma model to 3+1D, by taking into account the small x JIMWLK evolution of the color charge distributions \cite{Schenke:2016ksl,McDonald:2020oyf}, or by including sub-eikonal corrections due to the finite longitudinal thickness of the colliding nuclei based on colored particle-in-cell methods (CPIC) \cite{Gelfand:2016yho,Ipp:2017lho,Ipp:2020igo}.

In this work we develop a framework to perform 3+1D classical Yang-Mills simulations of the initial energy deposition in heavy-ion collisions, which as in \cite{Gelfand:2016yho,Ipp:2017lho,Ipp:2020igo} take into account the finite longitudinal extent of the colliding nuclei. While the colliding nuclei are still propagating on the light-cone, the collision region no longer remains a point but stretches to a diamond like region in space-time, such that the formation of the Glasma begins as soon as the two nuclei begin to overlaps. Similar to previous studies \cite{Gelfand:2016yho,Ipp:2017lho,Ipp:2020igo} this also entails the explicit propagation of the color currents of the colliding nuclei, which is in contrast to the boost-invariant case where they are considered to be static.
\\
Within a simple model of the color charge distribution of each nucleus, we perform a detailed investigation of the dynamics during and shortly after the collision as a function of the longitudinal thickness of the colliding nuclei, and contrast our results with the high-energy limit of infinitely thin shocks.
Subsequently, we develop a more physical model that connects the color charge distributions in the colliding nuclei to parton distributions inside the nuclei, and we discuss the rapidity profiles and fluctuations that emerge within this model.

Due to the computational complexity of $3+1$ D simulations, we will perform all of our simulations for the $SU(2)$ gauge group rather than the physical gauge group $SU(3)$ of QCD. Even though we do not expect to see any qualitative changes between the $SU(2)$ and $SU(3)$ dynamics, quantitative values of observables will change, and should therefore not be compared directly to experimental results.

This work is organized as follows: We first introduce the general formalism to discretize 3+1D Yang-Mills equation in \ref{Sect2}, followed by the initial condition for gauge fields and currents. In Section \ref{Sect3}, we present our results for 3+1D collision of individual color charges based on MV-model, followed by exploring rapidity dependence of the observable. We present our results for collision with realistic charge distributions in Section \ref{Sect4}, where we investigate the fluctuation at lower energies. 

\section{General formalism for 3+1D collision in Yang-Mills theory}\label{Sect2}
Based on the Color Glass Condensate (CGC) effective description of high-energy QCD, the initial state energy deposition and early time dynamics in high-energy heavy-ion collisions can be described semi-classically by solving classical Yang-Mills equations of motion for the gluon fields $A^{\mu}$ in the presence of fluctuating color charges $\rho$, which characterize the nuclear parton content. Even though a complete analytical treatment of the Yang-Mills equation is not possible, it is remarkable that the initial state immediately after the collision $(\tau=0^{+})$ can be determined analytically for boost-invariant collisions in the high-energy limit \cite{Lappi:2006fp,Kovner:1995ts}. Beyond the time of the collision $\tau=0$, where a far-from-equilibrium Glasma is produced \cite{Krasnitz:1998ns,Krasnitz:1999wc}, the classical Yang-Mills dynamics becomes highly non-linear, and additional analytic insights can only be obtained in the limit where one or both of the two sources are considered to be weak such that the equations of motion linearize.

Nevertheless, important insights into the early time non-equilibrium dynamics have been established based on numerical simulations \cite{Lappi:2003bi,Schenke:2012wb,Fukushima:2011nq,Romatschke:2006nk}, where in order to describe the non-linear dynamics of the boost invariant Glasma, the effectively 2+1 dimensional classical Yang-Mills equations are discretized on a lattice and solved numerically in the forward light cone i.e. for $\tau>0$. Below we explain how to generalize this setup to simulate the collision of nuclei with a finite longitudinal thickness in 3+1D collisions, where in contrast to the boost invariant high-energy limit, the entire space-time dynamics of the collision has to be simulated numerically, including the explicit evolution of color charges before, during and after the collision.
\\

\subsection{Discretization of gauge fields and currents}
We follow standard procedure in the context of classical-statistical lattice gauge theory simulations \cite{Kovner:1995ja,Krasnitz:1997zj} and start by discretizing the classical Yang-Mills Hamiltonian in temporal axial $A^{t}=0$ gauge on a three dimensional $N_{x}\times N_{y} \times N_{z}$ lattice with lattice spacing $a_\mu$ in the $\hat{\mu}$ direction  \cite{Krasnitz:1998ns,Lappi:2003bi}
\begin{align}\label{eq:Hamiltonian}
    H_{YM}={} & 
   \sum\limits_{x,I,J}\frac{\sqrt{-g}a^3}{g^2a_I^2a_J^2}(-g_{II})(-g_{JJ}) {\rm ReTr}\big[1-U_{IJ}(x)\big] \notag\\ 
  &{}+\sum\limits_{x,I}\frac{a_I^2}{g^2\sqrt{-g}a^3}(-g_{II})\frac{E^I_{a,x}E^I_{a,x}}{2}
\end{align}
 where $x$ denotes the lattice site, $a=1 \cdots N_{c}^2-1$ is the color index and $g_{\mu\nu}=(+1,-1,-1,-1)$ denotes the Minkowski metric. We will use $I,J=x,y,z$ to denote the spatial Lorentz indices,  $i,j=x,y$ to denote the transverse Lorentz indices, and $\alpha,\beta,\mu,\nu$ to denote the four dimensional Lorentz indices in Minkowski $(t,\mathbf{x_\perp},z)$ or light-cone $(x^{+},x^{-},\mathbf{x_\perp})$ coordinates, with $\mathbf{x_\perp}=(x,y)$ denoting the transverse coordinates. We note that within the Hamiltonian formulation in Eq.~\ref{eq:Hamiltonian}, the electric field strength is represented in terms of the lattice electric field variables 
 \begin{eqnarray}
 \label{eq:EFields}
    E^I_{a,x}=\frac{ga^3}{a_I}\sqrt{-g}(-g^{IJ})\partial_{0}A_{a,J}(x+\hat{I}/2+\hat{t}/2)
\end{eqnarray}
while the magnetic field strength is given in terms of the lattice plaquette variables
 \begin{align}
 \label{eq:Plaq}
 U_{x,IJ}={}&U_{x,I}U_{x+\hat{I},J}U^\dagger_{x+\hat{J},I}U^\dagger_{x,J}\notag \\
 &{}\simeq \exp\big(iga_I a_J F_{IJ}(x+\hat{I}/2+\hat{J}/2)\big)\;
\end{align}
which are formed of $\rm{SU(N_{c})}$ group valued lattice gauge links
\begin{eqnarray}
\label{eq:Links}
    U_{x,I}\simeq \exp\big(ig a_I A^I(x+\hat{I}/2)\big)\;.
\end{eqnarray}
and the pre-factors in the definitions in Eqns.~\ref{eq:EFields}, \ref{eq:Plaq} have been arranged such that all lattice variables are explicitly dimensionless. Within the CGC framework the incoming nuclei are described by eikonal color currents $J_{R/L}^\mu(x)$ which propagate along the light cones and provide a source
\begin{eqnarray} 
J^{\mu}(x)=J_R^\mu(x)+J_L^\mu(x)
\end{eqnarray}
for the classical gluon fields, where the subscripts $R$ and $L$ represent the nuclei coming in from the right and left respectively. Before the collision the initial conditions for the currents $J_{R/L}^\mu(x)$ in covariant $(\partial_{\mu} A^{\mu}=0)$ gauge are given in terms of the color charge densities $\rho^a_R(x^+,\bm{x_\perp})$ $\rho^a_L(x^-,\bm{x_\perp})$ of the two colliding nuclei as
\begin{eqnarray}
      J_R^\mu(x)=\delta^{\mu-}\rho^a_R(x^+,\bm{x_\perp})t^a \nonumber\\
      J_L^\mu(x)=\delta^{\mu+}\rho^a_L(x^-,\bm{x_\perp})t^a
\end{eqnarray}
where $t^a$ are generators in the fundamental representation. However, during the collision both currents $J_{R/L}^\mu(x)$ will receive a color-rotation, which in the 3+1 D setup has to be calculated by solving dynamical equations of motions for the currents $J_{R/L}^\mu(x)$. We therefore also discretize the color currents on the lattice, where by keeping track of the relevant light-cone ($\pm$) components, the currents are defined as 
\begin{eqnarray}
    J^{\pm}_{x,dyn}(t)={}&g a^{3} \frac{J^{0}_{R/L}(x+\hat{z}/2) \pm  J^{z}_{R/L}(x+\hat{z}/2)}{\sqrt{2}}\; \notag\\
    J^{\pm}_{x,stat}(t+\frac{a_t}{2})={}&g a^{3} \frac{J^{0}_{R/L}(x+\hat{t}/2) \pm  J^{z}_{R/L}(x+\hat{t}/2)}{\sqrt{2}} \notag \\
    &{}
\end{eqnarray}
with "dynamical" (dyn) and "static" (stat) currents on alternating half-integer time slices, as usual in a leap-frog scheme. 
\subsection{Evolution equations \& Gauss Law}
By performing the variation of the lattice Hamiltonian w.r.t to electric fields and gauge fields, one obtains the Hamiltonian equations of motion for the lattice gauge link and electric field variables. We employ a leap frog algorithm with time step $a_t=0.08\times min(a_z,a_\perp)$, where gauge links are defined at every full time step whereas the electric fields are calculated for every half-integer time step, such that the update rule for the lattice gauge links takes the form
\begin{align}
\label{eq:updateU}
 U_{I,x}(t+a_t)={}& \exp\Big(-i\frac{a_I^2 a_t}{\sqrt{-g}a^3}(-g_{II})E_{I,x}(t+\frac{a_t}{2}) \Big) U_{I,x}(t)\notag \\
 &{}
\end{align}
whereas for the evolution of the lattice electric fields one also has to take into account the coupling to the eikonal currents, such that the update rule for the lattice electric fields is given by
\begin{eqnarray}
\label{eq:updateEi}
  \lefteqn{  E^i_{a,x}(t+a_t/2)-E^i_{a,x}(t-a_t/2)=(-2\sqrt{-g}a^3a_t)}\notag\\
  &&\bigg(\frac{-g_{ii}}{a_i^2}\bigg)\sum_{j}\bigg(\frac{-g_{jj}}{a_j^2}\bigg){\rm ReTr}\big[it^a\big(U_{ij}(x)-U_{i-j}(x)\big)\big] \notag\\
  &&
\end{eqnarray}
\begin{eqnarray}
\label{eq:updateEz}
  \lefteqn{E^z_{a,x}(t+a_t/2)-E^z_{a,x}(t-a_t/2) =(-2\sqrt{-g}a^3a_t)}\notag \\ 
  &&\bigg(\frac{-g_{zz}}{a_z^2}\bigg)\sum\limits_{i}\bigg(\frac{-g_{ii}}{a_i^2}\bigg){\rm ReTr}\big[it^a\big(U_{zi}(x)-U_{z-i}(x)\big)\big]\notag \\ &&~~~~~~-a_t\frac{J_{a,x,dyn}^{+}(t)-J_{a,x,dyn}^{-}(t)}{\sqrt{2}a_z}\notag\\
  &&
\end{eqnarray}

Due to the explicit appearance of the currents on the rhs of Eq.~\ref{eq:updateEz}, the color currents $J^{\pm}_{x}(t)$ also have to be treated as dynamical degree of freedom as -- in contrast to the boost-invariant high-energy limit -- they are present not only on the infinitesimal boundary of the light-cone but throughout the entire simulation volume and thus affect the evolution of the classical Yang-Mills fields. Since in the eikonal limit the different components of the current are related by $J^{0}_{R/L}(x)=\mp  J^{z}_{R/L}(x)$, it is straightforward to construct the dynamical equation of motion of the currents from the (covariant) charge conservation equation $D_\mu J^\mu=0$ as $\partial_{0} J^{\pm}(x)=\mp D_{z}J^{\pm}(x)$, and we employ the following update rules
\begin{align}
\label{eq:updateJs}
    J_{x,dyn}^\pm(t+a_t) - J_{x,dyn}^\pm(t)={}& \mp a_t D_z^F J_{x,stat}^\pm(t+\frac{a_t}{2}) \notag \\ 
    J_{x,stat}^\pm(t+\frac{a_t}{2})- J_{x,stat}^\pm(t-\frac{a_t}{2})={}& \mp a_t D_{x,z}^B J_{x,dyn}^\pm(t)\notag
    \\
   &{} 
\end{align}
where $D^{F/B}_I$ denote the forward and backward covariant derivatives
\begin{align}
   D_{x,I}^FX={}&\big(U_{x,I}X_{(x+\hat{I}),I}U^\dagger_{x,I}-X_{x,I}\big)/a_I\; \notag\\    D_{x,I}^BX={}&\big(X_{x,I}-U^\dagger_{(x-\hat{I}),I}X_{(x-i),I}~U_{(x-\hat{I}),I}\big)/a_I \notag\\
   &{}
\end{align}
We note that due to the leap-frog discretization in Eqns.~\ref{eq:updateJs}, the dynamical currents do not propagate exactly at the speed of light, but instead satisfy the same lattice dispersion relation as the lattice gauge fields, which converges to a light-like dispersion in the continuum limit. This is different from CPIC method where the color charges have light-like dispersion and the dispersion of the gauge field depends upon the details of the numerical scheme \cite{Ipp:2018hai}. 
However, as in CPIC, the lattice version of the Gauss law constraint 
\begin{align}
\label{eq:GaussLaw}
    D_{x,I}^{B} E^I_{x}(t+\frac{a_t}{2})={}&\frac{1}{\sqrt{2}}\bigg(J^+_{stat}(t+\frac{a_t}{2})+J^-_{stat}(t+\frac{a_t}{2})\bigg)\notag \\
    &{}
\end{align}
is automatically satisfied at each time, as long as it is satisfied by the initial conditions, as can be checked straightforwardly by evaluating the time derivative of Eq.~\ref{eq:GaussLaw}, based on the equations of motion for the lattice gauge links, electric fields and currents as in Eqns.~\ref{eq:updateU}-\ref{eq:updateJs}.

\subsection{Initial Conditions for 3+1D collisions}
\begin{figure}%
    \centering
    \subfigure{{\includegraphics[width=0.3\textwidth]{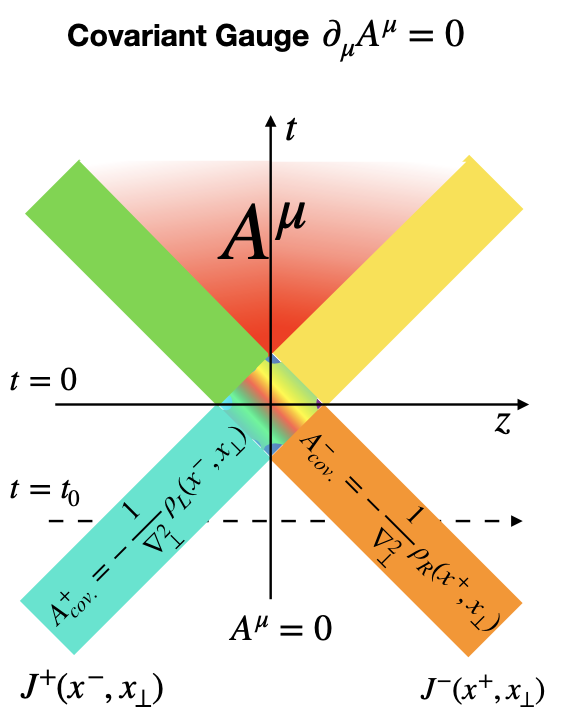}}}
    \subfigure{\includegraphics[width=0.3\textwidth]{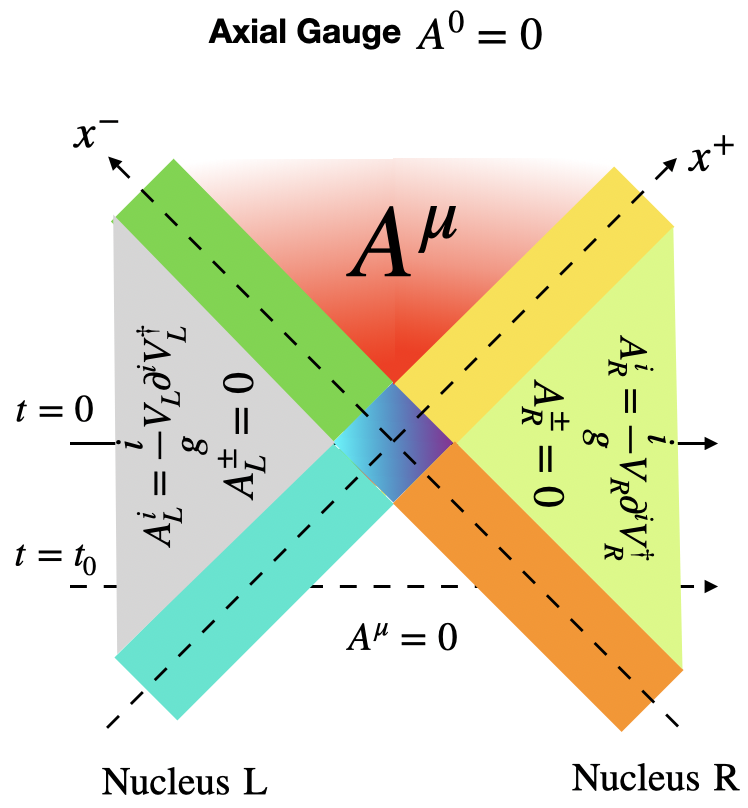}}
    \caption{Gauge fields for the two colliding nuclei with finite longitudinal extent $(R_\gamma\simeq\frac{R}{\gamma})$ in covariant gauge (top) and light-cone or temporal axial gauge (bottom).}
    \label{GaugeVsGauge}
\end{figure}

Since for 3+1 D collisions of extended nuclei, one has to simulate the dynamics of the color charges before, during and after the collision, the initial conditions for the above evolution equations have be formulated at Minkowski time $t_0<0$ before the collision, where the colliding nuclei are well separated from each other. Since the color charges inside the two nuclei do not interact with each other before the wave-packets overlap, the initial conditions are then determined by the superposition of the analytic solutions for the gauge fields in the presence of the individual color charges of the two nuclei, as illustrated in Fig.~\ref{GaugeVsGauge}. 

\begin{figure*}
    \centering
    \includegraphics[width=0.97\textwidth,height=7cm]{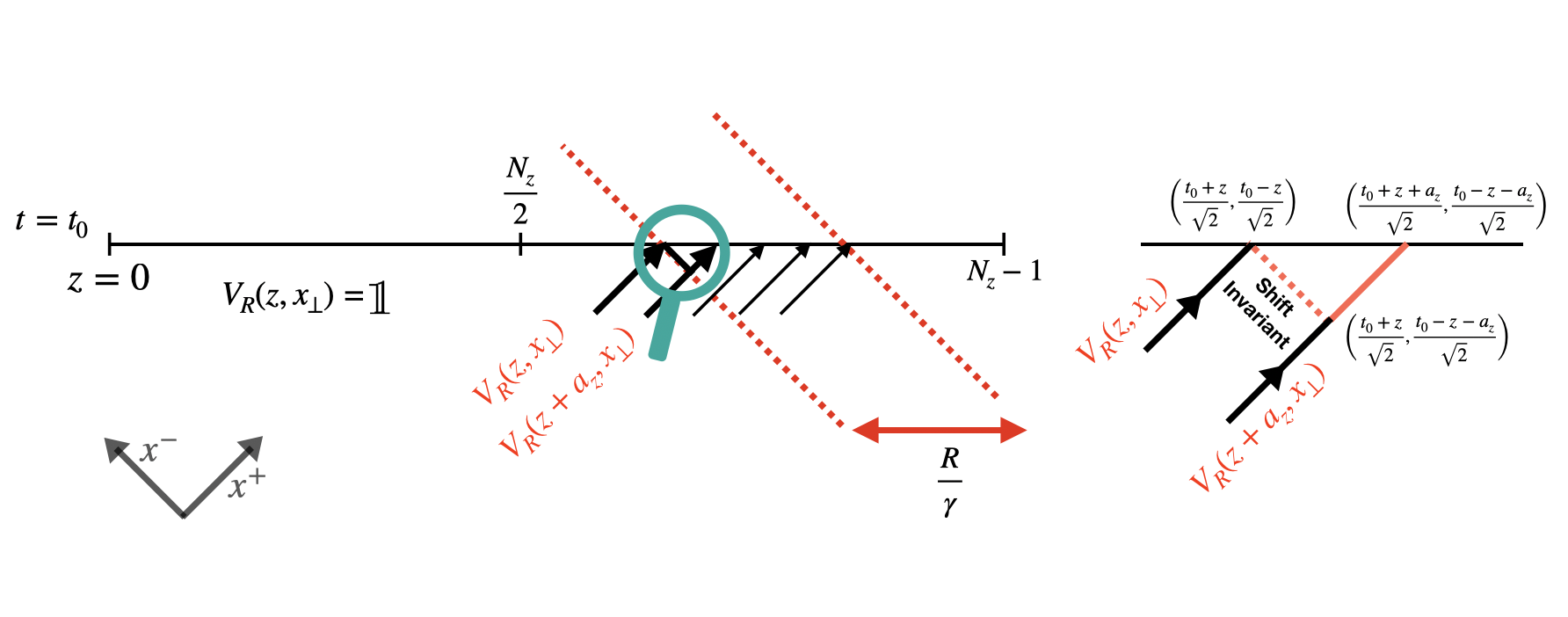}
    \caption{Illustration of light-like Wilson lines on t-z grid for the left moving nucleus with a finite longitudinal extent ($R_\gamma=\frac{R}{\gamma}$)}
    \label{HolyGrail}
\end{figure*}
Specifically, in the covariant $\partial_{\mu}A^{\mu}=0$ gauge, the solution to classical Yang-Mills equations before the collision takes the form
\begin{align}
\label{eq:CovGauge}
    A^\pm_{cov.}(x^\mp,\bm{x_\perp})={}&-\frac{1}{\nabla_\perp^2}\rho_{\rm{L,R}}(x^\mp,\bm{x_\perp}) \qquad A^i_{cov.}=0\notag\\
    &{}
\end{align}
where we have explicitly assumed that -- at the initial time -- the incoming nuclei are sufficiently far separated from each other, such that charge distributions  $\rho_{\rm{L,R}}(x^\mp,\bm{x_\perp})$ do not overlap with each other. 

We note that in $\partial_{\mu}A^{\mu}=0$ gauge the gauge potentials $A^{\pm}$, only have support in the vicinity of the two light-cones, where color charges are present, as seen in top panel of Fig.~\ref{GaugeVsGauge}. However, for real time lattice simulation it is convenient to employ the temporal axial $(A^0=0)$ gauge condition, and the corresponding initial conditions can be obtained by performing a gauge transformation, which eliminates the gauge potentials $A^{\pm}$ prior to the collision. By following previous works \cite{JalilianMarian:1996xn,Kovchegov:1996ty,Lappi:2007ku}, the corresponding gauge transformation can be expressed as
\begin{align}\label{eq:mainWils}
V(x)={}&V_R(x^+,\bm{x_\perp}) V_L(x^-,\bm{x_\perp})
\end{align}
where the light-like Wilson lines $V_{L/R}$ associated with the left and right moving nuclei are determined by
\begin{eqnarray}
    \partial_{+}V_R(x^+,\bm{x_\perp})&=&igA_{cov.}^-V_R(x^+,\bm{x_\perp}),\nonumber \\
    \partial_{-}V_L(x^-,\bm{x_\perp})&=&igA_{cov.}^{+}V_{L}(x^{-},\bm{x_\perp})\;
\end{eqnarray}
such that $V_{L/R}$ are given by the light-like Wilson lines
\begin{align}
\label{eq:wilsonline}
    V_R(x^+,\bm{x_\perp})={}&\mathcal{P}\exp\bigg(+ ig\int_{-\infty}^{x^+}dy^+A_{cov.}^-(y^+,\bm{x_\perp})\bigg),\notag \\
    V_L(x^-,\bm{x_\perp})={}&\mathcal{P}
    \exp\bigg(+ ig\int_{-\infty}^{x^-}dy^-A_{cov.}^{+}(y^{-},\bm{x_\perp})\bigg)
\end{align}
By performing the corresponding gauge transformation, the initial conditions in temporal axial gauge then take the form
\begin{eqnarray}
A^\pm=0 \qquad A^i=\frac{i}{g}V\partial^i V^\dagger
\end{eqnarray}
which is illustrated on the right panel of Fig.~\ref{GaugeVsGauge}. By initializing the simulation, at a sufficiently early time $t=t_0$, where the incoming nuclei are well separated from each other, one finds that at each particular point $z$ at most one of the two Wilson lines $V_{L/R}$ is different from the identity, and the Wilson lines in Eq.~\eqref{eq:mainWils} effectively commute. Similarly, at each particular point $z$ the corresponding gauge fields $A^{i}$ vanish, or reduce to the well known solutions $A^{i}_{L/R}$ for individual nuclei in the respective light-cone gauge \cite{JalilianMarian:1996xn,Kovchegov:1996ty,Lappi:2003bi}, as indicated in  Fig.~\ref{GaugeVsGauge}.

So far we have discussed the structure of the initial conditions in the continuum, and we will now address the corresponding lattice implementation. Starting from a given distribution of color charges $\rho_{L/R}(t_0,z,\bm{x_\perp})$ at initial time $t_0$ (as shown in Fig.~\ref{GaugeVsGauge}) discretized on a spatial $x,y,z$ grid, we first compute the covariant gauge $A^\pm_{cov.}(t_0,z,\bm{x_\perp})$ fields according to Eq.~\ref{eq:CovGauge}, and subsequently construct the discretized version $V_{L/R}(t_0,z,\bm{x_\perp})$ of the light-like Wilson lines. Since before the collision, the Wilson lines $V_{R}(x^{+},\bm{x_\perp})$ are independent of $x^{-}$, the light-like Wilson lines $V_{R}(t_0,z,\bm{x_\perp})$ can be defined to end on the lattice points $x,y,z$ at initial time $t_0$ as illustrated in the left panel of Fig.~\ref{HolyGrail}.

If we consider the left moving nucleus, which is initially located on the right hand side of the lattice, the corresponding Wilson line $V_{R}(t_0,z,\bm{x_\perp})$ is equal to the identity for all points $z$ which at the time $t_0$ are located to the left of the incoming color charges. Starting from $V_{R}(t_0,z=0,\bm{x_\perp})=\mathds{1}$ at the left boundary of the lattice $(z_{i}=0)$, the Wilson lines $V_{R}(t_0,z,\bm{x_\perp})$ for $z>0$ can be constructed successively based on the relation
\begin{eqnarray}\label{eq:GenWil}
\lefteqn{V_{R}(t_0,z+a_z,\bm{x_\perp})=}  \notag\\ &&\mathcal{P}\exp\bigg(+ ig\int_{x^{+}=\frac{t_0+z}{\sqrt{2}}}^{x^{+}=\frac{t_0+z+a_z}{\sqrt{2}}} dx^+A_{cov.}^-(x^+,\bm{x_\perp})\bigg) V_{R}(t_0,z,\bm{x_\perp}) \notag\\
&&
\end{eqnarray}
Exploiting again the invariance of the $A_{cov.}^-(x^+,\bm{x_\perp})$ gauge fields under shifts along the $x^{-}$ direction, the additional color rotation in Eq.~\ref{eq:GenWil} can then be approximated as
\begin{eqnarray}
 \lefteqn{\mathcal{P}\exp\bigg(+ ig\int_{x^{+}=\frac{t_0+z}{\sqrt{2}}}^{x^{+}=\frac{t_0+z+a_z}{\sqrt{2}}} dx^+A_{cov.}^-(x^+,\bm{x_\perp})\bigg)}\notag \\
 &&\simeq \exp\bigg(+ig\frac{a_z}{\sqrt{2}}  A_{cov.}^-\big(t_0,z,\bm{x_\perp}\big)\bigg)   
\end{eqnarray}
as indicated in the right panel of Fig.~\ref{HolyGrail}. Based on this procedure, the lattice discretized version of the Wilson lines for the right and left sitting nucleus is then given by
\begin{equation}
     V_{R,L}(t,\bm{x_\perp},z) = \prod_{z_i} \exp\bigg(\frac{\pm iga_z}{\sqrt{2}}A^\mp_{R,L}(t,\bm{x_\perp},z_i)\bigg)
\end{equation}
By applying the gauge transformation in Eq.~\ref{eq:Links} to the lattice gauge links, the initial conditions for the lattice gauge links are then  determined as
\begin{align}
    U_i(t,\vec{x})=V(\bm{x_\perp},z)V^\dagger(\bm{x_\perp}+\hat{i},z)\; \qquad     U_{z}(t,\vec{x})=\mathds{1}\;
\end{align}
Next, in order to initialize the lattice electric fields $E_{x,I}\big(t+\frac{a_t}{2}\big)$, we make use of the update rule for the gauge links in Eq.~\ref{eq:updateU}, to express
\begin{equation}\label{electricfield}
 E_{x,I}\big(t+\frac{a_t}{2}\big) = \frac{-i a^3}{a_I^2 a_t} {\rm log} \bigg( U_{x,I}(t+a_t)U_{x,I}^\dagger(t)\bigg)\;
 \end{equation}  
 which relates the electric fields to the gauge links at times $t_0$ and $t_0+a_t$. By following the same procedure as outlined above, the gauge links $U_{x,i}(t+a_t)$ are constructed from the color charges propagated by a single time step according to
\begin{align}
    \rho^a_{L,R}(t+a_t,\bm{x_\perp},z)={}&\frac{1}{N_z}\sum_{k_z}e^{\frac{2\pi ik_z z}{N_z}}\left\{\begin{matrix} e^{\mp i\omega a_t} :k_z<\frac{N_z}{2}\\[3pt] e^{\pm i\omega a_t} :k_z\geq \frac{N_z}{2} \end{matrix}\right\}\notag\\
    &{}\sum_{z'}e^{\frac{-2\pi ik_z z'}{N_z}}\rho^a_{L,R}(t,\bm{x_\perp},z')
\end{align}
with
\begin{equation*}
     \omega^2 = \frac{2}{a_z^2}\bigg(1-{\rm cos} \Big(\frac{2\pi k_z}{N_z}\Big)\bigg)\;,
\end{equation*}
such that the evolution of the color charges in the initialization step satisfies the lattice dispersion. 

While the above procedure provide initial conditions for the lattice gauge links and electric fields in temporal axial gauge, the color charges $\rho_{L/R}(t_0,\bm{x_\perp},z)$ are still given in covariant gauge. Instead of performing a gauge transformation of the charges, we exploit Gauss Law to determine the static currents as

\begin{equation}
\label{eq:InitialJstat}
    J_{stat}^\pm\big(t+\frac{a_t}{2},\bm{x_\perp},z\big)=\sqrt{2}\sum_{i=1,2} D_iE^i\big(t+\frac{a_t}{2},\bm{x_\perp},z\big)
\end{equation}
where the factor of $\sqrt{2}$ comes from the transformation between Minkowski and light-cone coordinates. Subsequently, the initial value of the dynamical currents $J_{dyn}$ is set by performing half a time step of evolution as
\begin{align}
&J_{dyn}^\pm(t+a_t,\bm{x_\perp},z)={}\pm\frac{a_z}{a_t}\frac{1}{N_z}\sum_{k_z}e^{\frac{2\pi ik_z z}{N_z}}\\
&{}\left\{\begin{matrix} 1 \hspace{1.6cm}:k_z=0\\ \bigg(\frac{e^{\pm i\omega  a_t}-1}{1-e^{\frac{-2\pi i k_z}{N_z}}}\bigg):k_z<\frac{N_z}{2}\\[2pt] 0\hspace{1.9cm}: k_z=\frac{N_z}{2}\\\bigg(\frac{e^{\mp i\omega a_t}-1}{1-e^{\frac{-2\pi i k_z}{N_z}}}\bigg) :k_z>\frac{N_z}{2} \end{matrix}\right\}\sum_{z'}e^{\frac{-2\pi ik_z z'}{N_z}}J_{stat}^\pm(t+\frac{a_t}{2},\bm{x_\perp},z') \nonumber
\end{align}
Starting from these initial conditions the subsequent dynamics, including the evolution in the diamond shaped collision region in Fig.~\ref{GaugeVsGauge}, is then simulated numerically.

\subsection{Observables}
Starting from the initial conditions outlined above, we simulate the dynamics of the collision in Minkowksi coordinates $x^{\mu}=(t,x,y,z)$ by solving the classical Yang-Mills equations for the lattice gauge links and electric fields, along with the evolution equations for the eikonal currents. While in principle the evolution can be performed up to arbitrary late times, in practice the incoming nuclei will approach to boundary of the lattice at some finite time after the collisions. Since we implement fixed boundary conditions along the longitudinal direction\footnote{Note that in temporal axial gauge, the gauge potentials are non-vanishing and different in the limit $z \to \pm \infty$ (c.f. Fig.~\ref{GaugeVsGauge}), such that periodic boundary conditions can not be used.} the color charges can not pass through the boundary and we have to stop our simulations before this occurs. 

When investigating the initial energy deposition during the collision and early-time dynamics of the Glasma, we will primarily focus on the evolution of energy momentum tensor $T^{\mu\nu}(x)$, which we compute as
\begin{align}
    T^{00}(x)={}&\frac{1}{2}\Big(E_{loc}^2(x)+B_{loc}^2(x)\Big)\nonumber \\
    T^{IJ}(x)={}&\frac{1}{2}\Big(E_{loc}^2(x)+B_{loc}^2(x)\Big)\delta^{IJ}- E_{loc}^{I,a}(x) B_{loc}^{J,a}(x) \notag \\
    T^{0I}(x)={}&\epsilon^{IJK}\Big(E^{J,a}_{loc}(x)\times B^{K,a}_{loc}(x)\bigg)
\end{align}
where $E_{loc}^2(x)=E_{loc}^{I,a}(x)E_{loc}^{I,a}(x)$, and $E_{loc}(x)$ and $B_{loc}(x)$ are local electric and magnetic fields calculated using smeared operator definitions, such that they are defined at same position i.e $(x+a_I/2+a_t/2)$ as
\begin{align}
    E^{a}_{I,loc}(x)={}&\frac{a_I}{2 a^3}\bigg(E^a_{I}(x)+U ^{\dagger}_{I}(x-\hat{I})E^a_{I}(x-\hat{I})U_{I}(x-\hat{I})\bigg) \notag \\
    B^{I,a}_{loc}(x)={}& \epsilon^{IJK} \frac{a_I}{4 a^3}~\textrm{ReTr} \bigg(it^a\Big( U_{JK}(x)+ U_{J -K}(x)\notag\\
    &{}+U_{-J-K}(x) +U_{-JK}(x)\Big)\bigg)\;.
\end{align}

Besides the space-time evolution of the energy momentum tensor, we will also consider the evolution of the field intensity of longitudinal $(\|)$ and transverse $(\perp)$ components of the (chromo-) electromagnetic fields 
\begin{align}
   B_{\|}^2(x)=\frac{2}{a_x^2a_y^2} \textrm{ReTr}\big[\mathds{1}-U_{xy}(x)\big]\;\notag\\
    B_{\perp}^2(x)=\sum_{i=x,y}\frac{2}{a_i^2a_z^2}\textrm{ReTr}\big[\mathds{1}-U_{iz}(x)\big]\;\notag\\
    E_{\|}^2(x)=\frac{1}{2a_x^2a_y^2}\sum_{a=1}^{N_c^2-1}\big[E_{z}^{a}(x)\big]^{2}\notag\\
    E_{\perp}^2(x)=\sum_{i=x,y}\frac{1}{2a_i^2a_z^2}\sum_{a=1}^{N_c^2-1}\big[E_{i}^{a}(x)\big]^2
\end{align}
Since the focus of our investigation is to understand the time evolution and the longitudinal structure of the fields, we will frequently perform averages over the transverse, which we denote as
\begin{eqnarray}
\bigg\langle T^{00}(t,z) \bigg\rangle_{\perp} = \frac{1}{N_xN_y} \sum_{\bm{x_{\perp}}} T^{00}(t,\bm{x_\perp},z)   
\end{eqnarray}

\section{3+1D Collisions of individual color charges}
\label{Sect3}
Based on the above simulation framework, we will now study the initial energy deposition and early time dynamics of the Glasma. Before we turn to simulations involving realistic models of the color charge distributions of the colliding nuclei, it proves insightful to first consider the collision of individual ensembles of color charges, to test the framework and develop an intuitive picture of the underlying dynamics.

We follow the previous works \cite{Krasnitz:1998ns,McLerran:1993ni}, and sample the transverse distribution of color charges based on the McLerran-Venugopalan (MV) model as
\begin{align}
\label{eq:2Dcharge}
    \rho^{a~(2D)}_{L/R}(x,y)={}&\sum_{k_x,k_y}e^{2\pi i \Big(\frac{k_xx}{N_x}+\frac{k_yy}{N_y}\Big)}\frac{k^2}{k^2+m^2}e^{-k^2/2\Lambda^2}\notag\\
    &{}\sum_{x',y'}Q_s\sqrt{a_xa_y}\zeta^{a}(x',y')e^{-2\pi i \Big(\frac{k_xx'}{N_x}+\frac{k_yy'}{N_y}\Big)}
\end{align}
where $\zeta(x,y)$ are Gaussian random numbers, $m \sim \Lambda_{QCD}$ is an infrared regulator enforcing color neutrality on large distance scales, $\Lambda$ is an ultra-violet cut-off suppressing color charge fluctuations on short distance scales  \footnote{We note that in the absence of a cut-off scales $\Lambda$, the color charge distribution is regularized by the lattice cut-off $\pi/a$. Introducing a cut-off scale at the level of the charge distribution renders the problem UV finite, and thus simplifies the realization of the continuum limit in our lattice calculations}, such that the model has a well defined continuum limit and if not stated otherwise we employ $m/Q_s=1$ and $\Lambda/Q_s=5$ in our simulations. 

Subsequently, the three dimensional color charge distribution $\rho^a_{L/R}(x,y,z)=\rho^{a~(2D)}_{L/R}(x,y) a_zT(z)$ is obtained by multiplying the transverse color charge distribution with the same Gaussian profile at each point
\begin{eqnarray}\label{eq:gaussprofile}
    T(z)=\frac{1}{\sqrt{2\pi R_\gamma^2}}e^{-z^2/2R_\gamma^2}
\end{eqnarray}
where the dimensionful parameter $R_\gamma=R/\gamma$ controls the longitudinal extent of the nucleus.

Since the initial color charge distributions are characterized in terms of the dimensionful scales $Q_s$ and $R_\gamma$, the latter can be used to set the scale of the lattice calculation by specifying the value of $Q_{s} a_\perp$ and $R_\gamma/a_z$. Generally, the (transverse) lattice spacing has to be chosen sufficiently small to avoid discretization errors $Q_{s} a_\perp \ll 1$, while at the same time the transverse simulation volume $N_{\perp} a_{\perp}$ should be large compared to the color charge correlation length $\sim 1/m$. Similarly, the longitudinal color charge distribution has to be smooth on the scale of a single lattice spacing $a_z \ll R_\gamma$, while at the same time the longitudinal extend of the lattice $N_{z} a_z$ has to be sufficiently large to allow for a long enough time evolution after the collision. Within the above setup, we have varied both the lattice spacings $Q_{s} a_\perp$, $R_\gamma/a_z$ as well as the lattice length $Q_{s} N_{\perp} a_\perp$ and $R_\gamma/N_za_z$ to check that discretization errors do not play a significant role, and if not stated otherwise we will present results for $N_\perp=128$, $N_{z}=2048$ with $Q_{s} a_\perp=0.125$ and $R_\gamma/a_z=16$ in the following.
\subsection{Stable propagation of color charges before and after the collision}
\begin{figure}
    \centering
    \subfigure{\includegraphics[width=0.37\textwidth]{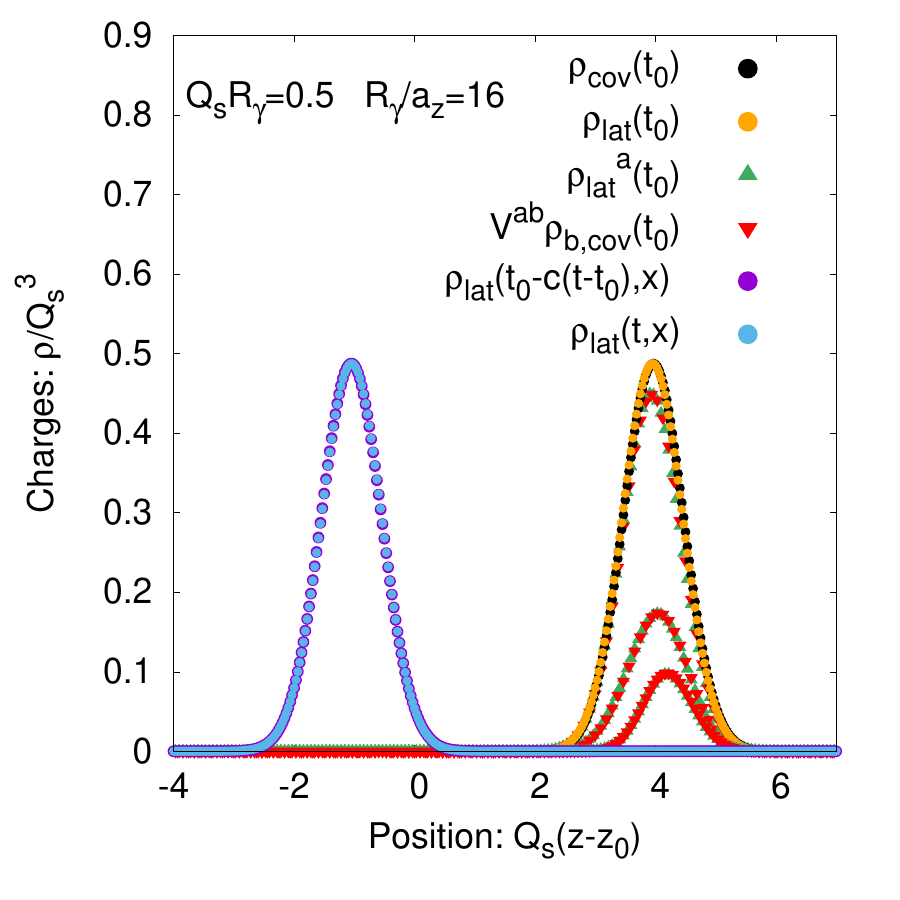}}
    \subfigure{{\includegraphics[width=0.38\textwidth]{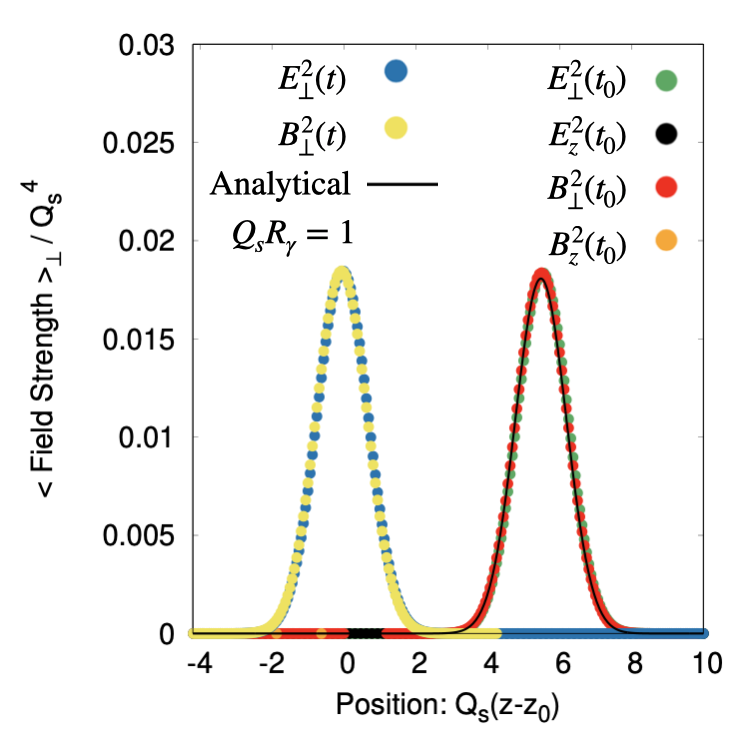}}}%
    \caption{Longitudinal profile of the color charges along with the different color components at initial time $t_0$ and after propagation for a time $t-t_0=5/Q_s$ (top). Electric and magnetic field strengths of the left moving nucleus along with the analytically obtained result at initial time $t_0$ and at some later time $t-t_0=5/Q_s$, prior to the collision (bottom).
}%
    \label{BeforeCollision}%
\end{figure}

Before we address the dynamics of the collision, we briefly verify that -- within our numerical setup -- the color charges of the individual nuclei propagate in a stable fashion. We illustrate this behavior in Fig.~\ref{BeforeCollision}, where the top panel shows the evolution of the longitudinal profiles of the color charge distribution $\rho(t,x,y,z)=\sqrt{\rho_{a}(t,x,y,z)\rho_{a}(t,x,y,z)}$ at a randomly chosen point $x,y$ in the transverse plane. By comparing the initial color charge distribution $\rho_{a}(t_0,x,y,z)$ determined according to Eq.~\ref{eq:InitialJstat}, to the
charge density $\rho_{a}(t_0,x,y,z)= V_{ab}(t_0,x,y,z)\rho_{b}^{\rm cov.}(t_0,x,y,z)$ re-constructed from the color charge distribution in covariant gauge, we observe an excellent agreement demonstrating that the re-construction of the charge density based on Gauss's law works as expected. By comparing the dynamically evolved charge distribution $\rho_{a}(t,x,y,z)$ to the translated initial conditions $\rho_{a}(t_0,x,y,z-c(t-t_0))$, we can further confirm that for sufficiently small lattice spacing $a_z \ll R$, the numerical dispersion of the currents is small, such that over the relevant time scales the nuclei propagate in a stable fashion at almost the speed the light. 
Beside the stable propagation of the color charges, it is also important that the gluon fields induced by the color charges propagate in a stable way along side the charges, as can be seen from the bottom panel of Fig.~\ref{BeforeCollision}, where we show the evolution of the longitudinal profiles of the average electric  and magnetic field strengths $\langle E_{\perp}^{2}(t,z) \rangle_{\perp}$, $\langle B_{\perp}^{2}(t,z) \rangle_{\perp}$, $\langle E_{\|}^{2}(t,z) \rangle_{\perp}$ and $\langle B_{\|}^{2}(t,z) \rangle_{\perp}$. By comparing the longitudinal field strength profiles in Fig.~\ref{BeforeCollision} at different times, one again concludes the nuclei propagate in a stable fashion over the relevant time scales; moreover the analytic result for the evolution of the field strength is given as: 
\begin{equation}
    E_\bot^2(t_0,z)=\frac{Q_s^2}{4}(N_c^2-1) T(z)^2\int \frac{d\mathbf{k_\perp}}{2\pi}\frac{1}{\mathbf{k_\perp}}\Big(\frac{\mathbf{k_\perp}^2}{\mathbf{k_\perp}^2+m^2}\Big)^2 e^{\frac{-\mathbf{k_\perp}^2}{\lambda^2}}
\end{equation}
which follows directly from Eqns.~\ref{eq:CovGauge} and \ref{eq:2Dcharge}, is well reproduced by our real-time lattice simulations, indicating the residual discretization errors are indeed small.

\begin{figure}
    \centering
    \subfigure{\includegraphics[width=0.38\textwidth]{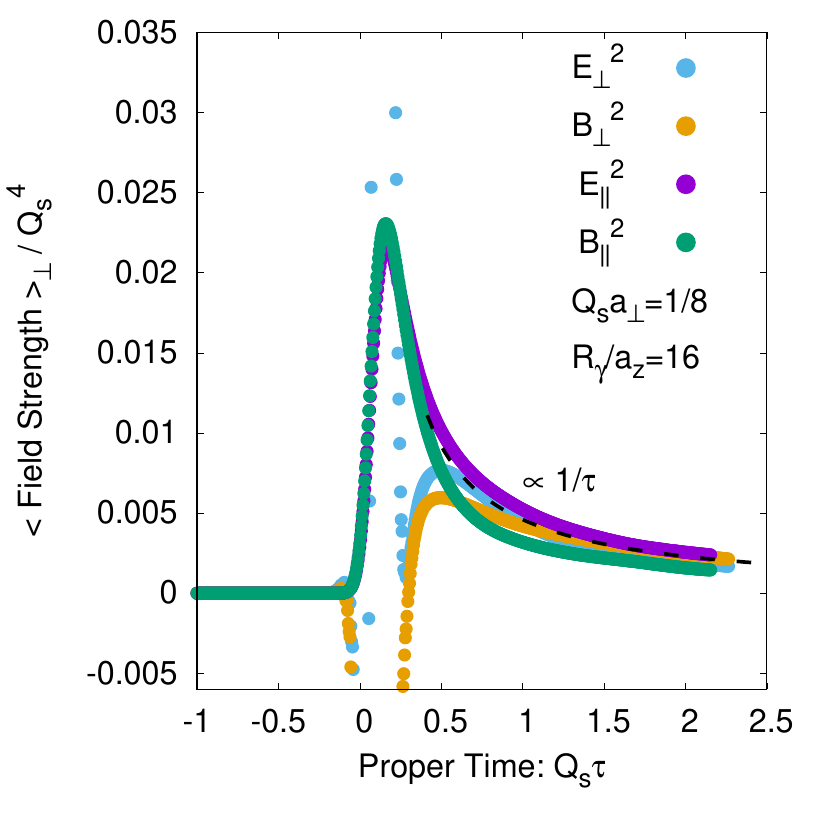}}\hfil
    \subfigure{\includegraphics[width=0.38\textwidth]{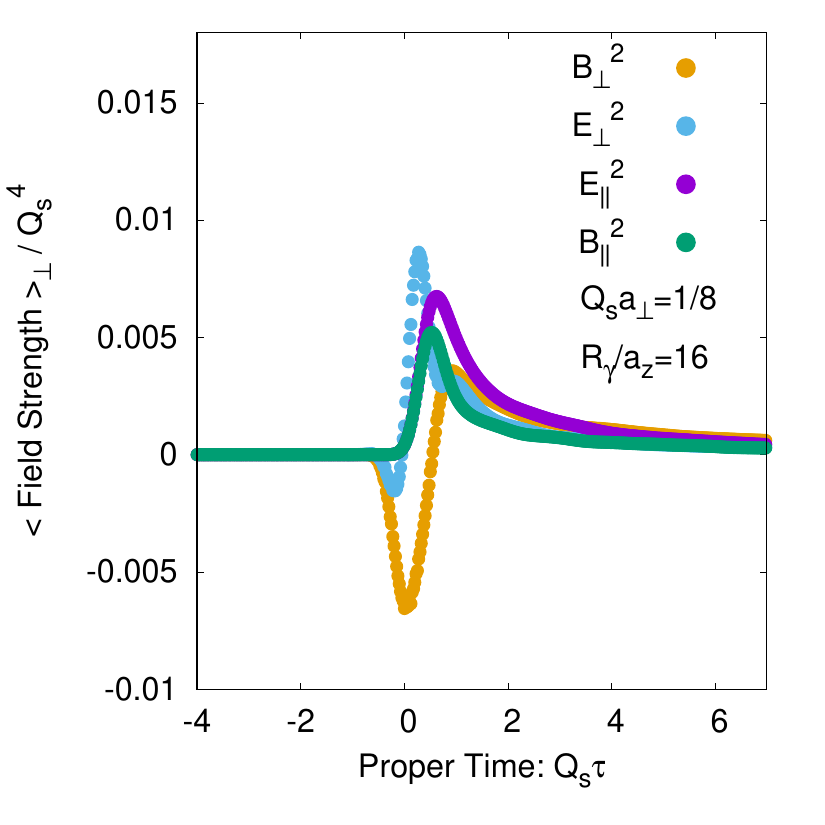}}
    \caption{Evolution of longitudinal and transverse components of the  electric and magnetic fields at the center of the collision for different longitudinal thickness: $Q_sR_\gamma=0.0625$ (top) and $Q_sR_\gamma=0.5$ (bottom).}
    \label{ThickVsThin}
\end{figure}
\begin{figure}[bt]
    \centering
    \subfigure{\includegraphics[width=0.385\textwidth]{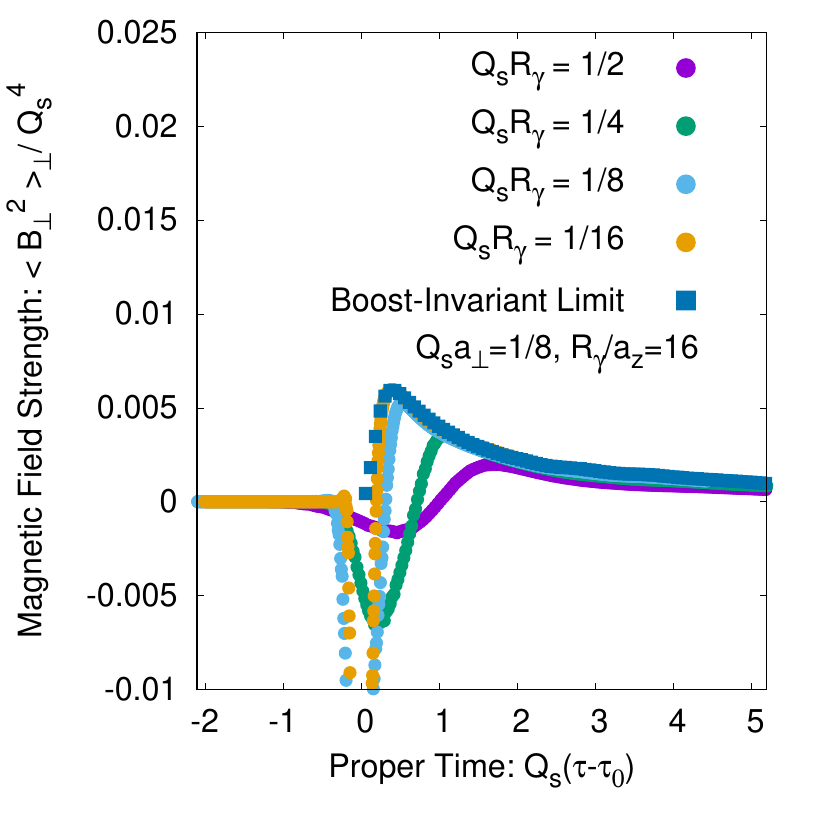}}\hfil
    \subfigure{\includegraphics[width=0.3885\textwidth]{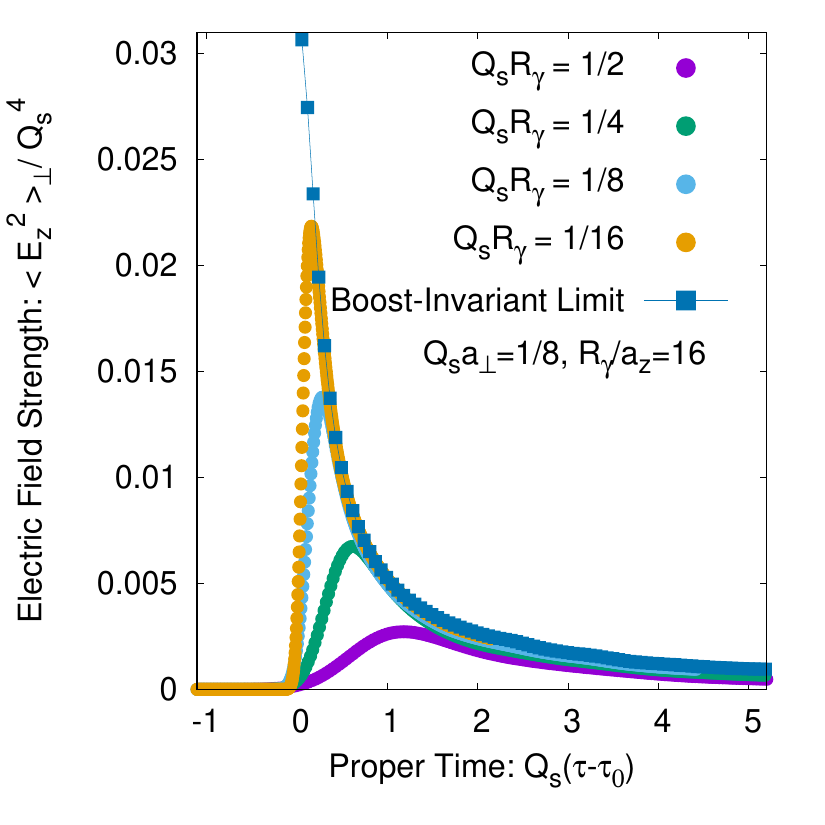}}
    \caption{Evolution of transverse magnetic (top) and longitudinal electric (bottom) fields at the center of the collision for different longitudinal thickness $Q_sR_\gamma$ along with the boost invariant limit.}
    \label{AfterCollision}
\end{figure}
\subsection{Evolution of the fields during and after the collision}
Now that we have established the validity of our setup, we will analyze the energy deposition and early time dynamics of the collisions. Before we present our numerical results, we briefly recall the structure of the Glasma fields in the high-energy boost invariant limit \cite{Kovner:1995ja,Lappi:2006fp,Kovner:1995ts} which will serve as a basis for comparison. Before the collision, the incoming nuclei feature the transverse electric and magnetic fields known as the Weiszaecker-Williams fields (WW)
\begin{align}
E^{i}_{WW}=\partial^0A^i_{R/L}\qquad
B^{i}_{WW}=\epsilon^{ij}\partial^zA^j_{R/L}
\end{align}
localized in narrow strips along the light-cones. Even though the structure of the fields during the collisions is not analytically accessible, it is well established that the initial state immediately after the collision $(\tau=0^{+})$ features boost invariant longitudinal electric and magnetic fields in the forward light-cone
\begin{align}
    \label{eq:initialglasmafields}
    E^{\eta}(\tau=0,\bm{x_\perp})={}&-ig \delta^{ij}
    V^\dagger(\bm{x_\perp})[A^i_{L}(\bm{x_\perp}),A^j_{R}(\bm{x_\perp})]V(\bm{x_\perp})\; \notag \\ B^{\eta}(\tau=0,\bm{x_\perp})={}&-ig \epsilon^{ij}
    V^\dagger(\bm{x_\perp})[A^i_{L}(\bm{x_\perp}),A^j_{R}(\bm{x_\perp})]V(\bm{x_\perp})\;  
\end{align}
where $V(\bm{x_\perp})$ is defined in Eq.~\ref{eq:mainWils} and we adapted the usual $\tau=\sqrt{t^2-z^2}$ and $\eta=\tanh^{-1}(z/t)$ coordinates, with longitudinal electric and magnetic fields in the Milne coordinates defined as $E^\eta=\frac{1}{\tau} F_{\tau\eta}$ and $B^\eta=-\frac{1}{2}\epsilon^{ij}F_{ij}$, which at mid-rapidity $(\eta=0)$ are equivalent to the fields in Minkowski coordinates i.e $E^\eta|_{\eta=0}=E^z$,$B^\eta|_{\eta=0}=B^z$. Subsequently, for $\tau>0$ the initial Glasma flux tubes in Eq.~\ref{eq:initialglasmafields} begin to expand in transverse space and loose their coherence, such that after a time scale $\sim 1/Q_s$ the Glasma features longitudinal and transverse color fields of comparable magnitude, resulting in a state of approximately vanishing longitudinal pressure $p_{L}=T^{\eta}_{~\eta}\simeq 0$ \cite{Lappi:2006fp}. While the evolution of the boost-invariant Glasma produced in the forward light-cone has been explored to detailed extent within numerical simulations \cite{Lappi:2006fp,Schenke:2012wb,Romatschke:2006nk} and (semi-)analytic calculations \cite{Kovner:1995ja}, we further note that also in the boost-invariant high-energy limit the eikonal charges receive a color rotation during the collision, and the transverse electric and magnetic fields continue to exist in narrow strips along the light-cones.

Beyond the high-energy boost invariant limit, the formation of the Glasma begins as soon as the color charge distributions of the incoming nuclei start to overlap and persists over an extended period of time until the colliding nuclei have passed through each other. Now in order to analyze the formation of the 3+1D Glasma, we first consider the evolution of the fields at the center of the collision ($z=0$), where one has $\tau=t$ and $\eta=0$ such that the descriptions in Minkowski $(t,z)$ coordinates and Milne $(\tau,\eta)$ coordinates coincide. We present our results in Fig.~\ref{ThickVsThin}, where we show the time evolution of the longitudinal and transverse electric and magnetic fields strength during and after the collision for two different values of $Q_sR_\gamma=1/4$ and $Q_sR_\gamma=1/16$ corresponding to different longitudinal thickness of the colliding nuclei. Since we want to focus on the creation of the Glasma, we have subtracted the field strength associated with Weiszaecker-Williams fields of the colliding charges, i.e. we consider $E^{2}_{\rm Glasma}(t,z)=E^{2}(t,z)-E^{2}_{WW}(t,z)$, and we have defined the origin of the coordinate system such that at $t=0$ the charge distributions of the colliding nuclei maximally overlap with each other. During the collision longitudinal electric and magnetic fields build up monotonically as the two nuclei pass through each other, while the transverse electric and magnetic components experience rapid changes. By the time that the incoming nuclei have passed each other, which corresponds to $Q_{s}t\simeq 0.25$ in the top panel and $Q_{s}t \simeq 1$ in the bottom panel, the transverse electric and magnetic fields become small again and longitudinal electric and magnetic fields dominate the energy density. While close to the boost-invariant limit for $Q_sR_\gamma=1/16$ the longitudinal electric and magnetic fields have approximately the same magnitude, the longitudinal magnetic field strength is suppressed compared to the longitudinal electric one away from the boost-invariant limit for $Q_sR_\gamma=1/4$. Eventually, as the colliding nuclei have passed through each other, transverse electric and magnetic fields are regenerated from their longitudinal counterparts, until at late times the different components become of comparable magnitude. Similar to the boost-invariant case, the different field intensities at the center of the collision $(z=\eta=0)$ then start to decay as approximately $\propto 1/\tau$ as indicated by the black dashed lines in Fig.~\ref{ThickVsThin}

\begin{figure*}
    \centering
    {\includegraphics[height=14cm,width=0.9\textwidth]{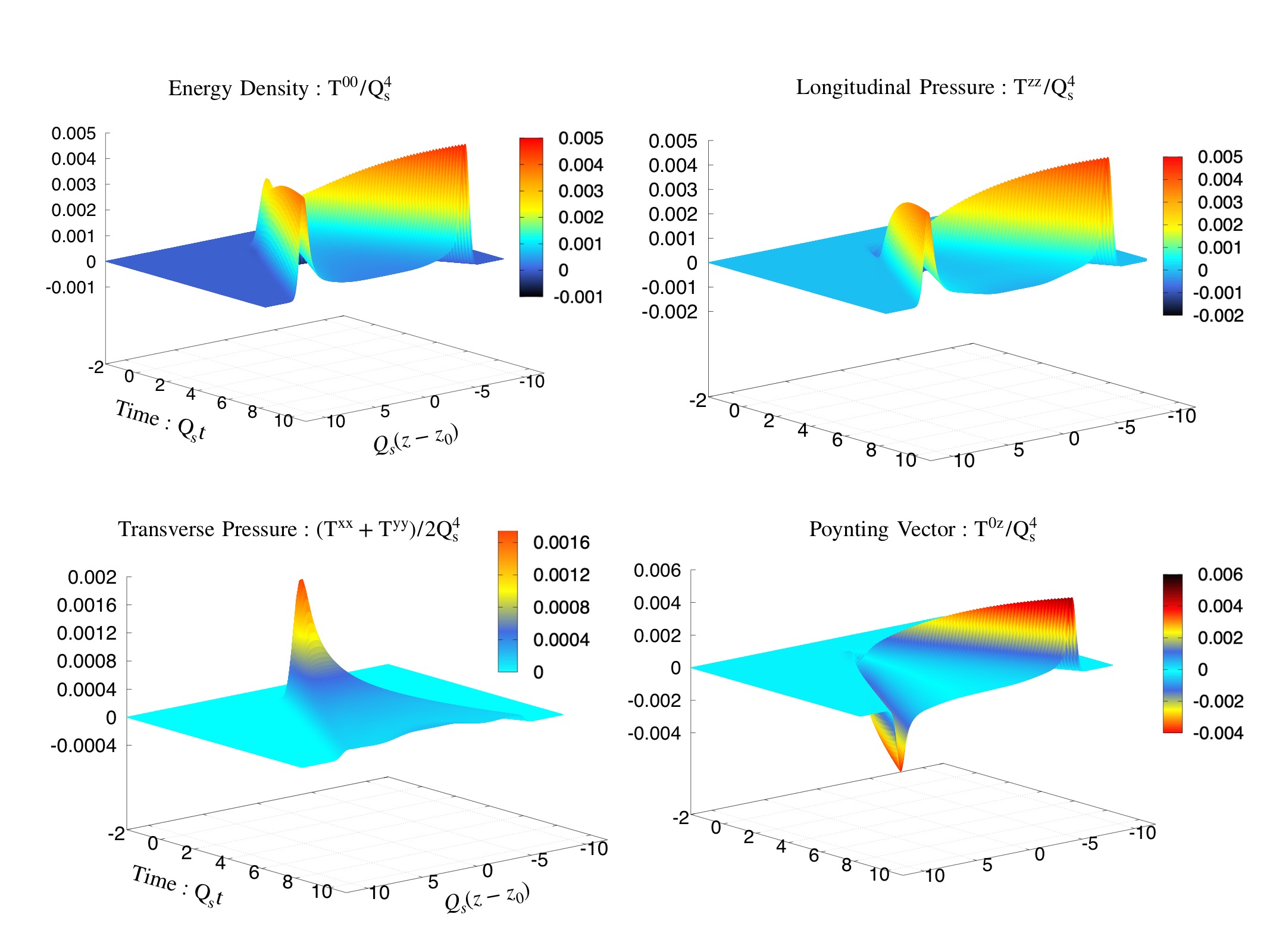} }\hfil%
    \caption{Space-time evolution of  different components of the stress energy tensor for the thick nuclei $(Q_sR_\gamma=0.5)$}%
    \label{StressComp}%
\end{figure*}

Next we will investigate the dependence of the initial energy deposition and early time dynamics on the longitudinal thickness. In Fig.~\ref{AfterCollision} we present the evolution of transverse magnetic and longitudinal electric fields for different values of $Q_sR_\gamma$ characterizing the longitudinal thickness of the colliding nuclei.\footnote{Similar behavior can be observed for the longitudinal magnetic and transverse electric components which are not depicted here.} Alongside the results from 3+1D numerical simulations, we also show results for the boost invariant limit, obtained by performing $2+1$D boost-invariant classical Yang-Mills simulations for the same color charge distributions, using the infinite Wilson lines $V_{L/R}(x_{\perp})$ as an input. Starting from the collision of thick nuclei with $Q_sR_\gamma=1/2$, where the collision takes a significant amount of time and the evolution of $B_{\perp}^2(t,z=0)$ and $E_{\|}^2(t,z=0)$ shows a smooth transition between the different stages, the decrease/increase of the transverse magnetic/longitudinal electric field strength during the collision sharpens significantly as the collision becomes shorter and shorter for smaller values of $Q_sR_\gamma$. Conversely, the evolution of the fields after the collision for $Q_s t \gtrsim Q_sR_\gamma$ is rather insensitive to the longitudinal thickness, and the results for 3+1D collisions smoothly approach the boost-invariant result as $Q_sR_\gamma \to 0$.

So far we have focused on the time evolution in the center of the collision ($z=0$), and we will now analyze space-time evolution of longitudinal profiles of the collision in more detail. Instead of showing results for the individual field strength components, we will focus on the evolution of the dominant components $T^{00},T^{0z},T^{zz}$ and $T^{ii}$ of the energy momentum tensor, and similar to our previous result subtract the contributions $T^{00}_{WW}=T^{zz}_{WW}=\pm T^{0z}_{WW}$ of the Weiszaecker-Williams fields to the energy momentum tensor. By performing three independent simulations, where in the first case we simulate the full collision, while the other two simulations simply propagate of the left/right moving charges, the subtraction takes into account the non-zero dispersion of the color charges due to residual discretization errors, and the energy-momentum tensor of the Glasma $T^{\mu\nu}_{\rm Glasma}=T^{\mu\nu}_{full} -T^{\mu\nu}_{WW,L}-T^{\mu\nu}_{WW,R}$ vanishes identically before the two nuclei collide. 

Our results for the space-time evolution evolution of the energy-momentum tensor are compactly summarized in Fig.~\ref{StressComp}, where the different panels show the $t,z$ dependence of the various components of $\langle T^{\mu\nu}_{\rm Glasma} \rangle_{\perp}$ in the lab-frame averaged over the transverse plane. By focusing on the evolution of the transverse pressure $(T^{xx}+T^{yy})/2$, one clearly observes the energy deposition in the central region where $(T^{xx}+T^{yy})/2$ increases during the collision, exhibits a pronounced peak and subsequently decreases due to the rapid longitudinal expansion of the Glasma. However, in addition to the energy deposition in the central region, we also observe rather large contributions to $T^{00},T^{0z}$ and $T^{zz}$ in the vicinity of the two light-cones. While it is intuitively clear that the non-equilibrium plasma produced away from central region should feature sizeable velocities in the longitudinal direction and therefore contribute significantly to $T^{00},T^{0z}$ and $T^{zz}$ in the lab-frame, the magnitude of contributions is surprisingly large compared to the transverse pressure. Even though we can not clearly rule out that these contributions may arise due to artifacts of the lattice discretization, we have checked explicitly that the observed behavior remains unchanged when we decrease the lattice spacing, as discussed in more detail in Appendix \ref{Appendix:Continnumlimit}. Since to the best of our knowledge such behavior has not been reported previously in the context of 3+1D Glasma simulations, clarifying the detailed structure of the fields in the vicinity of the light-cone will require further numerical and analytical investigations in future.

\begin{figure}
    \centering
    \subfigure{{\includegraphics[width=0.4\textwidth]{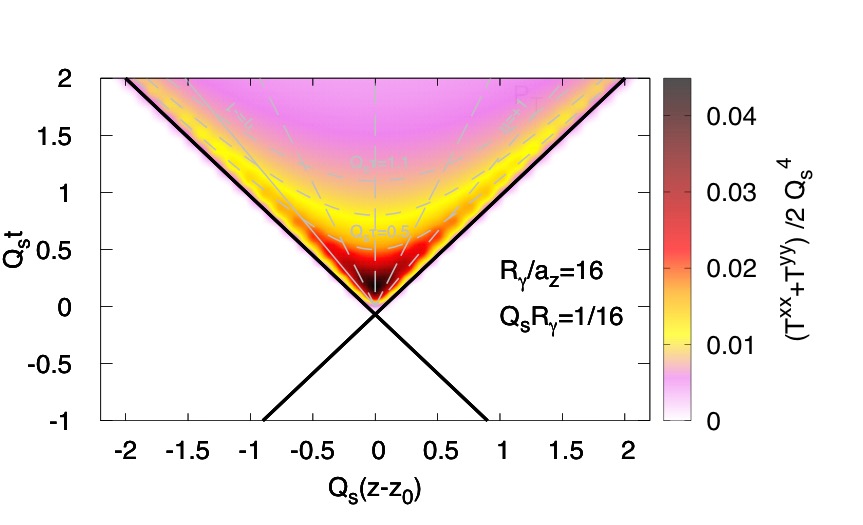}}}
    \subfigure{{\includegraphics[width=0.4\textwidth]{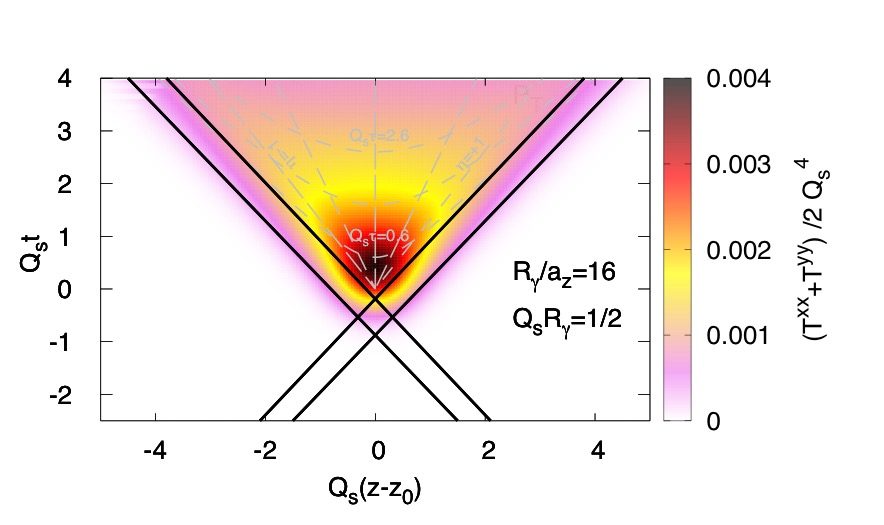}}}
    \caption{Heat-map plots for the space-time evolution of the transverse pressure $\langle P_T(t,z)\rangle_{\perp}$ overlayed with grey dashed $\tau,\eta$ lines for thin nuclei $Q_sR_\gamma=0.0625$ (top) and thick nuclei $Q_sR_\gamma=0.5$ (bottom).}%
    \label{heatmap}%
\end{figure}

While the results in Fig.~\ref{StressComp} were obtained for the collision of rather thick nuclei ($Q_sR_\gamma=1/2$), it is also interesting to investigate how the space-time profiles change when varying the longitudinal thickness $Q_sR_\gamma$ of the colliding nuclei. We investigate this behavior in Fig. \ref{heatmap}, where we present heat-map plots of the space-time evolution of the transverse pressure $(T^{xx}+T^{yy})/2$ for $Q_sR_\gamma=1/2$, overlayed with the $\tau,\eta$ coordinate system in the forward light-cone, by indicating lines of constant proper time $Q_s \tau$ and constant space-time rapidity $\eta$.
\begin{figure}
    \centering
    \includegraphics[width=0.42\textwidth]{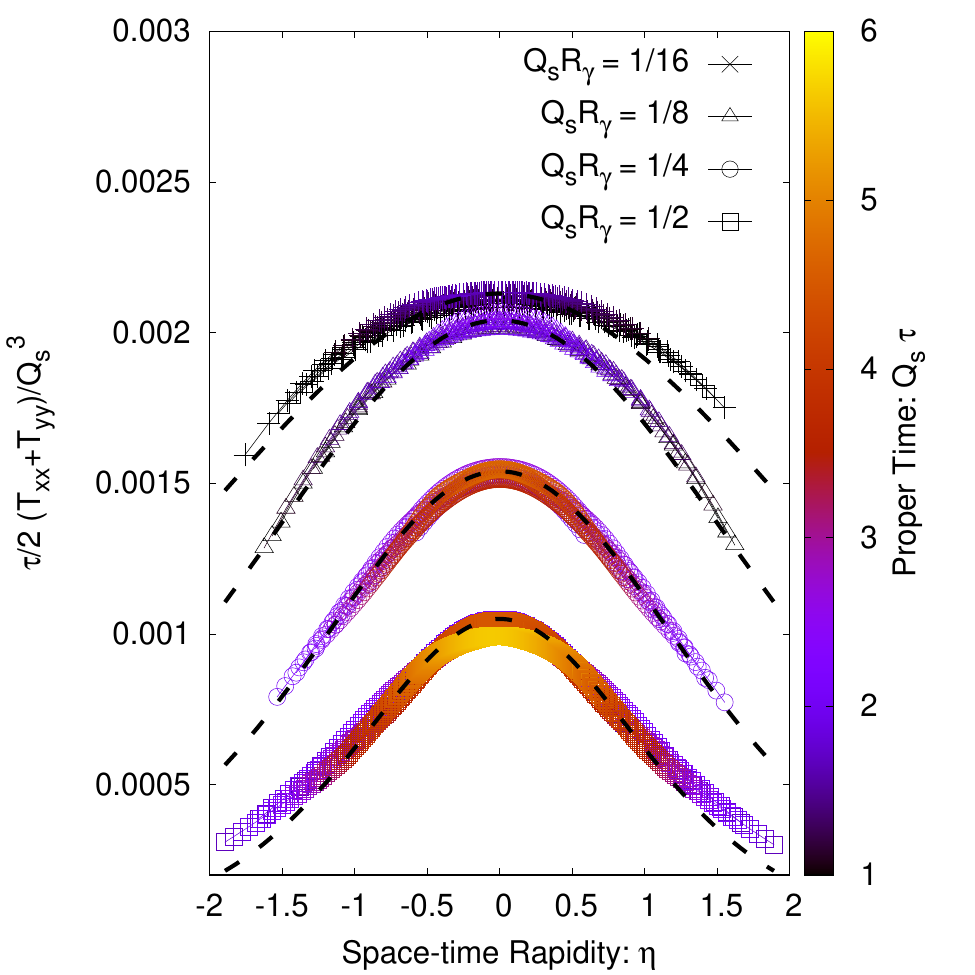}
    \caption{Rapidity profiles of the transverse pressure for various nuclear thickness $Q_sR_\gamma$. Color coding shows the different proper-time.}%
    \label{Observables}%
\end{figure}
While the results for the collision of thin nuclei ($Q_sR_\gamma=1/16$) closely resemble the behavior in the boost invariant limit, as lines of constant transverse pressure $(T^{xx}+T^{yy})/2$ coincide well with lines of proper time $Q_s \tau$ within the central rapidity region ($\eta \lesssim 1$), clear deviations from boost-invariance emerge when considering the collision of nuclei with a large longitudinal extent ($Q_sR_\gamma=1/2$). Most importantly, one observes that for $Q_sR_\gamma=1/2$ the transverse pressure is significantly reduced towards the edges of the forward light-cone, indicating a non-trivial space-time rapidity profile around mid-rapidity.

\subsection{Space-time rapidity profiles}\label{Subsection.3}

So far, we have leveraged our framework to study the space-time picture of the collision in Minkowksi space. Now we will look at the  non-trivial rapidity dependence of the observables, which arises naturally by including the longitudinal thickness of the colliding nuclei. Even though the mapping of $(t,z)$ data into $(\tau,\eta)$ coordinates is in principle straightforward, the availability of information on a discrete $t,z$ grid \footnote{Even though the time step $a_t$ is typically small, we only create output of the energy momentum tensor every 100 steps, in order to keep the overall data size reasonable.}  poses additional challenges, as a straightforward interpolation between data points can become problematic in the vicinity of the light cones. Due to these technical difficulties, we will re-strict ourselves to an investigation of the rapidity range $\eta \in (-1.25,1.25)$, and show the corresponding result for different $\tau$ as a function of $\eta$. We further note that the conversion of Minkowski $(t,z)$ space results to Milne coordinates $\tau,\eta$ can in fact be quite sensitive to the definition of the origin of the coordinate system, and we will always fix the origin $t=0,z=0$ at the space-time point, where the center of mass of the two nuclei coincides.

In Fig.~\ref{Observables}, we present the evolution of transverse pressure $\tau(T^{xx}+T^{yy})/2$ as a function of $\eta$ for different values of $Q_sR_\gamma$. Different color codings in Fig.~\ref{Observables} correspond to the results obtained at different proper times $\tau$, and the scaling of the transverse pressure by a factor of $\tau$ has been chosen such that -- beyond time scales $\tau \sim 1/Q_{s}$ -- the quantity $\tau(T^{xx}+T^{yy})/2$ shown in Fig.~\ref{Observables} becomes independent of proper time $\tau$ in the boost-invariant high-energy limit. Starting from the collision of thin nuclei with $Q_sR_\gamma=1/16$,  we observe the emergence of a boost-invariant plateau for $\eta \in (-0.8,0.8)$,  as already seen in the bottom panel of Fig.~\ref{heatmap}, where shortly after the collision the contours of constant transverse pressure follows the line of constant proper time $\tau$. When increasing the longitudinal thickness of the colliding nuclei, the transverse pressure of the Glasma created around mid-rapidity decreases and we see how the profiles are no longer flat around the central rapidity even for the late times $Q_st\gg Q_sR_\gamma$. Empirically, we find that the resulting rapidity profiles can be reasonably well described by the following functional form, which is indicated by the black dashed lines in Fig.~\ref{Observables},
\begin{eqnarray}\label{eq:FitFunction}
    \tau P_T(\eta)=\frac{\tau P_T(\eta=0)}{\rm{cosh}(\eta/\eta_R)}\;,
\end{eqnarray}
where $\tau P_T(\eta=0)$ is the pressure at mid-rapidity and $\eta_R$ describes the rapidity width. By looking at the extracted values of $\eta_R$ and $\tau P_T(\eta=0)$ in Tab.~\ref{Table1}, one observes that the rapidity width $\eta_{R}$ exhibits a strong dependence on the width $Q_sR_\gamma$ of the colliding nuclei, whereas the transverse pressure $\tau P_T(\eta=0)$ of the Glasma at mid-rapidity only decreases slowly with increasing thickness of the colliding nuclei, as can also be seen directly from Fig.~\ref{Observables}.
\begin{table}
\begin{tabular}{|l|l|l|}
\hline
~$\bm{Q_sR_\gamma}$      & ~~$\bm{\eta_{R}}$  & $\bm{\tau P_{T}/Q_s^3(\eta=0)}$  \\ \hline
~1/16~     &~2.48 ~~& ~~~~0.0021~ \\
  ~1/8~   &~1.69 ~~  & ~~~~0.0020~\\
  ~1/4~ & ~1.20~~ & ~~~~0.0015~\\
  ~1/2~ & ~0.98 ~~ & ~~~~0.0010~\\
  \hline
\end{tabular}
\caption{Parameters of fit function defined in Eqn. \ref{eq:FitFunction}}
\label{Table1}
\end{table}

When analyzing the rapidity dependence of the other components of the energy-momentum tensor, it is convenient to switch to the local rest frame (LRF), defined by the condition that $u^{\mu}_{LRF}$ is a time-like eigenvector of the energy-momentum tensor  $T^{\mu}_{~\nu}u^{\nu}_{LRF}=\epsilon_{LRF}u^{\mu}_{LRF}$. By diagonalizing the average
stress-energy tensor $T^\mu_{~\nu,Glasma}$ of the Glasma,\footnote{Note that, as discussed previously, the energy-momentum tensor of the Glasma $T^{\mu \nu}_{Glasma}$ is obtained by subtracting the contributions of the Weiszaecker-Williams fields of the colliding nuclei, prior to the diagonalization procedure} one gets the energy density and longitudinal pressure $P_L$ in this frame as
\begin{align}
\epsilon_{LRF}={}&\frac{1}{2}\bigg(T^{00}-T^{zz}+\sqrt{\big(T^{00}+T^{zz}\big)^2-4T^{0z}T^{0z}}\bigg)\notag\\
P_{L_{LRF}}=&{}\frac{1}{2}\bigg(T^{00}-T^{zz}-\sqrt{\big(T^{00}+T^{zz}\big)^2-4T^{0z}T^{0z}}\bigg)
\end{align}
\begin{figure}
    \centering
    \subfigure{{\includegraphics[width=7.5cm,height=6.2cm]{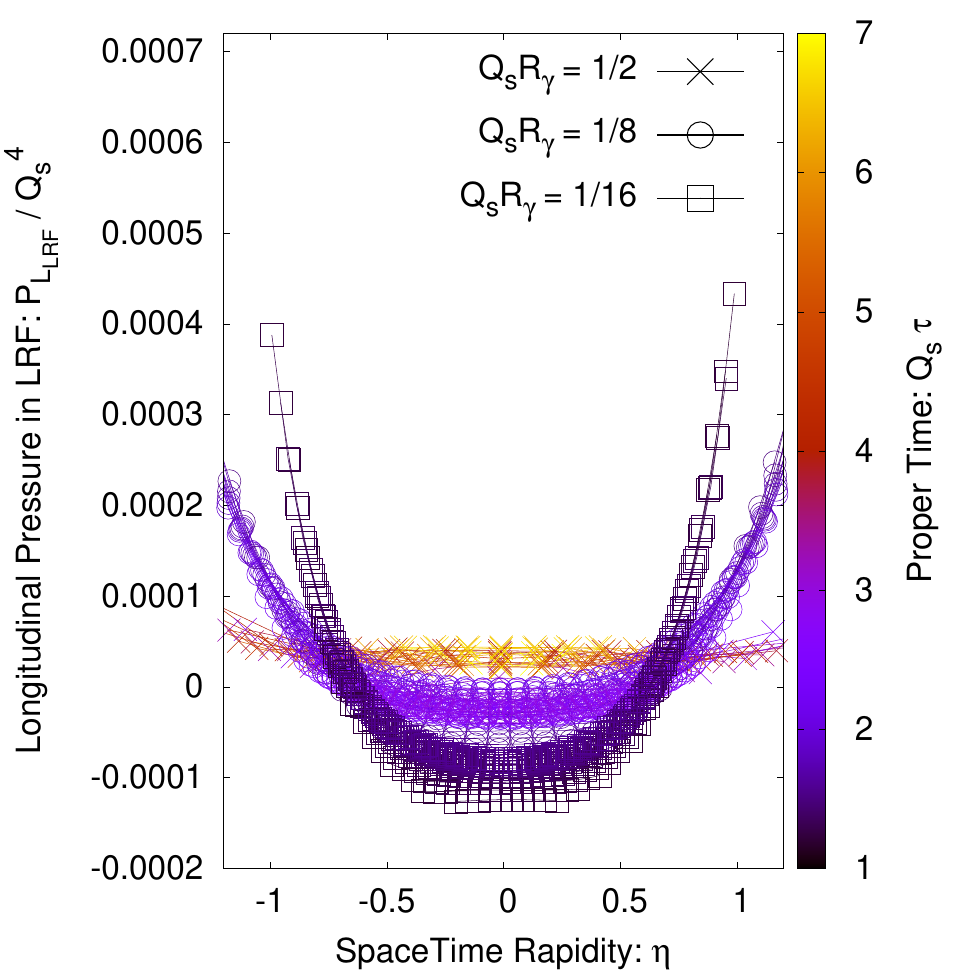}}}
    \subfigure{{\includegraphics[width=7.5cm,height=6.2cm]{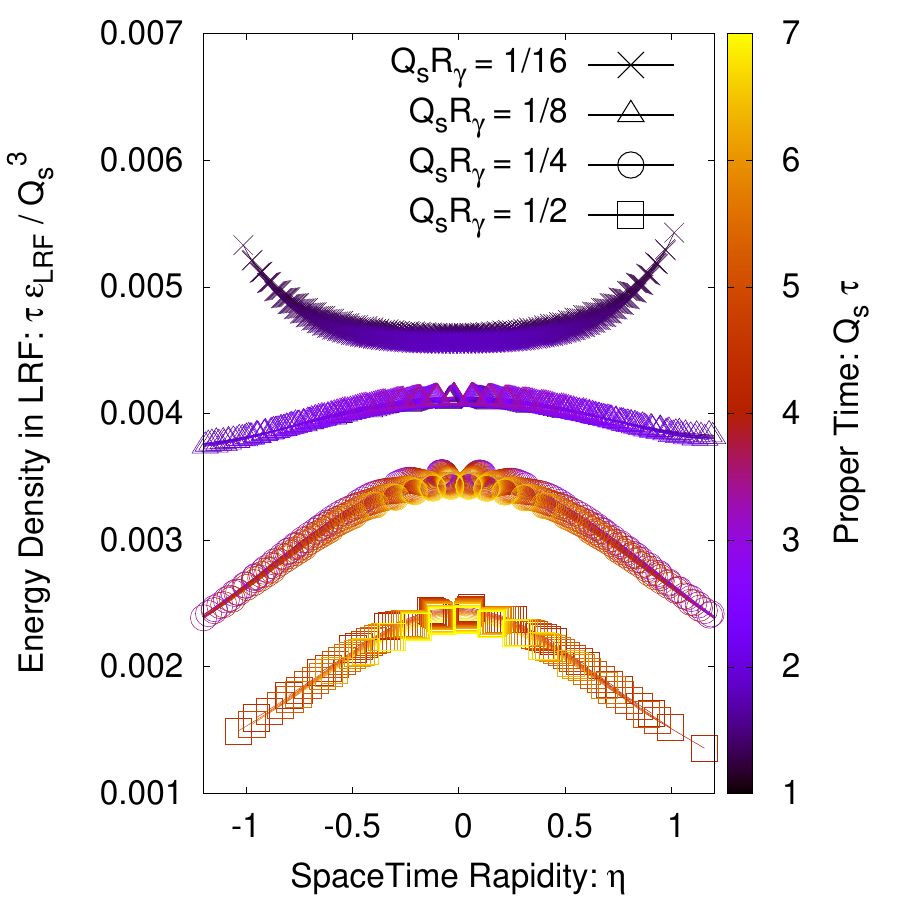} }}
    \caption{Rapidity profiles of the longitudinal pressure (top) and the energy density (bottom) in local-rest frame for different thickness of the colliding nuclei.}%
    \label{EigenValues}%
\end{figure}
We show our results in Fig.~\ref{EigenValues}, where we present results for the energy density $\tau~\epsilon_{LRF}$ and longitudinal pressure $P_{L_{LRF}}$ for different values of $Q_sR_\gamma$. Starting with the evolution of the longitudinal pressure depicted in the top panel, we find that for collisions of thick nuclei, the longitudinal pressure almost vanishes as the two nuclei have passed through each other, as seen for $Q_sR_\gamma=1/2$ at late times. With decreasing thickness, the longitudinal pressure starts out from negative values around mid-rapidity, and subsequently relaxes towards zero, in qualitive agreement with the well established behavior in the high-energy boost-invariant limit \cite{Lappi:2006fp,Kovner:1995ts}. Conversely, the rise of the longitudinal pressure $P_{L_{LRF}}$ at larger rapidities signifies clear deviations from boost invariance, and can be attributed to the spurious presence of fields on/near the light-cones in Fig.~\ref{StressComp}

In the bottom panel of Fig.~\ref{EigenValues} we present energy density for different longitudinal extent of nuclei, at late times where longitudinal pressure almost vanishes (top panel), such that $\epsilon_{LRF}\simeq 2P_T$ and $\tau \epsilon_{LRF} \simeq const$ is approximately constant. Similar to Fig.~\ref{Observables}, we again notice the emergence of a boost-invariant plateau in the high-energy limit $(Q_sR_\gamma=1/16)$, whereas for the collision of thick nuclei at lower energies $(Q_sR_\gamma=1/2,1/4)$ we see contrasting result which again signifies broken boost-invariance.

\section{3+1 D Collisions with realistic color charge distributions}\label{Sect4}

So far we have considered a simplistic model of color charge distributions inside each nucleus, to perform a detailed investigation of the longitudinal dynamics of the Glasma during and shortly after the collision. Evidently, to connect these simulations to realistic heavy-ion collisions, it is necessary to develop a more physical model of the color charge distributions, which reflect both the longitudinal and transverse structure of the colliding nuclei. Similar to the discussion in the boost invariant high energy limit \cite{Schenke:2012wb,McLerran:1993ni,McLerran:1993ka}, the basic idea of our construction will be to connect the color charge distributions $\rho_{L/R}(x^{\pm},\bm{x_\perp})$, to measurements of hadronic structure functions from the deep-inelastic scattering experiments. Below we develop a model of the three-dimensional structure of the color charge distribution based on the small-x transverse momentum distribution (TMDs), and subsequently perform simulations within this framework to study the effect of longitudinal fluctuations of the color charge distributions.

\subsection{Connection to small-x TMDs}
Generally speaking, the three-dimensional parton of nucleons and nuclei is encoded in an underlying Wigner distribution \cite{Belitsky:2003nz,Ji:2003ak} that contains information on the position and momenta of single partons inside a nucleon or nucleus. By disregarding position or momentum information, the Wigner distribution reduces to a transverse momentum parton distribution (TMD) or respectively to a generalized parton distribution (GPD). Conversely, if both position and momentum information are discarded, one obtains the standard collinear parton distribution function (PDF).

Even though a modeling of the color charge distribution based on the Wigner function would be desirable, little is known about this fundamental object, and we will therefore consider a parametrization of the color charge densities based on TMDs, with the three dimensional spatial structure of nucleons and nuclei imposed by hand according to a Monte Carlo Glauber model. Specifically, we will assume that the position and momentum dependence of the color charge distribution inside a nucleon can be factorized as
\begin{eqnarray}\label{eq:Ansatz}
\Big\langle \rho^a(x)\rho^b(y)\Big\rangle=\delta^{ab}T\Big(\frac{x+y}{2}\Big)\Gamma(x-y)\;,
\end{eqnarray}
where ${T}\left( \frac{x+y}{2}\right)$ captures the spatial structure of the colliding nucleon, and thus varies on length scales $\sim R_{p}$ and $\sim R_{p}/\gamma$, where $R_{p}$ is the proton radius, whereas the Fourier transform of $\Gamma\left(x-y\right)$ describes the transverse and longitudinal momentum dependence of color charges inside the nucleus, such that e.g. in the transverse plane $\Gamma\left(x-y\right)$ typically varies on distance scales $\sim 1/Q_s$.

Now in order to constrain the behavior of $\Gamma\left(x-y\right)$, we consider the operator definition of the dipole gluon TMD for a left moving nucleus~\cite{Petreska:2018cbf,Dominguez:2011wm}
\begin{align}
    x_2G^{(2)}(x_2,\mathbf{k_\bot})={}&\frac{4}{\langle p_A|p_A\rangle}\int_{-\infty}^{\infty}d\xi^+ d\xi'^+\frac{d^2\xi_\bot d^2\xi'_\bot}{(2\pi)^3}\notag\\ &{}e^{ix_2p^-(\xi^+-\xi'^+)}e^{-i\mathbf{k_\bot}(\xi_\bot-\xi'_\bot)}\notag\\
     &{} \Big\langle p_A| \textrm{Tr}\big[F^{i-}_{\xi}U_{[\xi,\xi']}F^{i-}_{\xi'}U_{[\xi',\xi]}\big]|p_A\Big \rangle
\end{align}
where $x_2$ is the longitudinal momentum fraction and $\mathbf{k_\bot}$ is the transverse momentum of the gluons. The gauge links $U_{[\xi,\xi']}$ and $U_{[\xi',\xi]}$ connecting the points $\xi$ and $\xi'$ ensures a gauge invariant definition of the TMD distributions.

Within the Color Glass Condensate effective theory, the matrix element $\langle p_A|...|p_A \rangle/\langle p_A|p_A\rangle$, is replaced by an average $\langle.\rangle$ over the color charge distribution \cite{Petreska:2018cbf,Dominguez:2011wm}
\begin{align}
    \label{eq:tmd}
     x_2G^{(2)}(x_2,\mathbf{k_\bot})={}&4\int_{-\infty}^{\infty}d\xi^+ d\xi'^+\frac{d^2\xi_\bot d^2\xi'_\bot}{(2\pi)^3}\notag\\ &{}e^{ix_2p^-(\xi^+-\xi'^+)}e^{-i\mathbf{k_\bot}(\xi_\bot-\xi'_\bot)}\notag\\
     &{} \Big\langle \textrm{Tr}\big[F^{i-}_{\xi}U_{[\xi,\xi']}F^{i-}_{\xi'}U_{[\xi',\xi]}\big]\Big \rangle
\end{align}
Based on eqn.~(\ref{eq:Links},\ref{eq:CovGauge}) the non-abelian field strength tensor $F^{i-}(\xi)$ and gauge links $U_{[\xi,\xi']},U_{[\xi',\xi]}$, can be calculated as functional of the color charge density $\rho$, such that the $x_2$ and $\mathbf{k_\bot}$ dependence of the dipole TMD $x_2G^{(2)}(x_2,\mathbf{k_\bot})$ is in fact entirely determined by specifying the $n-$point correlation functions of the color charge density. 

Evidently, the general relation between the correlation functions of $\rho$ and the dipole TMD is rather complicated, and we will thus simplify the problem by considering Gaussian correlations of color charges in the dilute limit, where the expression in Eq.~(\ref{eq:tmd}), can be expanded to lowest non-trivial order in $\rho$'s.
By evaluating the two-point correlation functions according to Eq.~(\ref{eq:Ansatz})
using average and difference coordinates, we then obtain
\begin{align}\label{eq:tmdresult}
x_2G^{(2)}(x_2,\mathbf{k_\bot}) ={}&\frac{4N_c C_{F}}{(2\pi)^3} \frac{\tilde{\Gamma}(\mathbf{k_\bot},x_{2}p^{-})}{\mathbf{k_\bot}^2} S_\perp
\end{align}
where $p^{-}=\sqrt{s_{NN}/2}$ is the large light-component of momentum of the nucleon in the lab frame and $S_\perp=\int d\xi^{+} \int d^2\xi_\perp T(\xi_\perp,\xi^{+})$ is the transverse area of the nucleon. Based on Eq.~(\ref{eq:tmdresult}), where $\tilde{\Gamma}$ denotes the Fourier transform of $\Gamma(x-y)$, one then concludes that the $\mathbf{k_\bot}$ dependence of the $x_2G^{(2)}(x_2,\mathbf{k_\bot})$ determines the transverse structure of the correlation function $\Gamma(x-y)$, whereas the longitudinal structure of the correlation function $\Gamma(x-y)$ is related to the $x_2$ dependence of the TMD.
\begin{figure*}%
    \centering
    \includegraphics[width=0.98\textwidth,height=7.5cm]{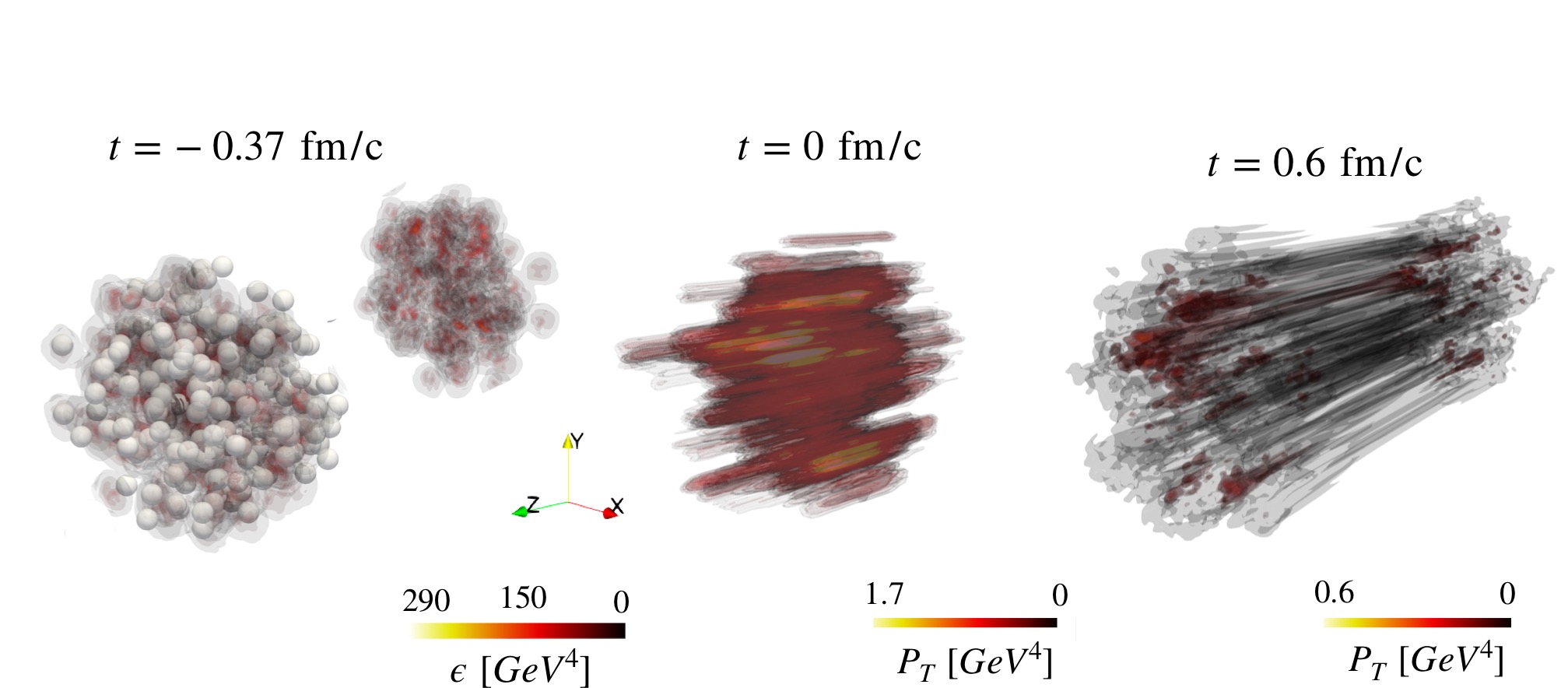}
    \caption{Illustration of 3D energy density with the position of nucleons as indicated by small grey spheres (left), transverse pressure during the collision (center) and after the collision (right) for a single event of $Au-Au$ collision at 200 GeV.}%
    \label{3dRenderingsForEvolution}%
\end{figure*}

Now in order to employ Eq.~(\ref{eq:tmdresult}) to determine the correlation function, we still need a parametrization of the gluon TMD $x_2G^{(2)}(x_2,\mathbf{k_\bot})$ as input to the calculation. We will employ the GBW Model~\cite{GolecBiernat:1998js}  --  a simple phenomenological model that has been fit to small-$x$ deep-inelastic scattering data -- where 
\begin{align}\label{gbwresult}
    x_2 G^{(2)}(x_2,\mathbf{k_\bot})\Big|_{x_2=x_0}={}&\frac{N_cS_\perp}{2\pi^3\alpha_s}\frac{\mathbf{k_\bot}^2}{Q^2_s(x_2)} \exp\Bigg[-\frac{\mathbf{k_\bot}^2}{Q_s^2(x_2)}\Bigg]
\end{align}
with the saturation scale $Q_{s}(x)$ parameterized as \cite{GolecBiernat:1998js}
\begin{eqnarray}\label{sat}
    Q_s(x)=Q_0x^{-\lambda}(1-x)\;,
\end{eqnarray}
with $\lambda=0.144$ and the value of $Q_0$ being set to $0.5$ GeV.
Since the color charges are assumed to be $x^{-}$ independent, the light-cone component $k^{+}$ vanishes identically for each source, such that in terms of the spatial momenta $k^{-}=-\sqrt{2}k_z$ and the momentum fraction is given by $x_2=\frac{k^{-}}{p^{-}}$. Based on Eq.~(\ref{gbwresult}), the correlation function $\tilde{\Gamma}(\mathbf{k_\bot},k_z)$ of the initial color charges is then obtained as
\begin{align}\label{correlation}
    \tilde{\Gamma}(\mathbf{k_\bot},k_z)=&\frac{8\pi}{g^2}\frac{N_c}{N_c^2-1}\frac{\mathbf{k_\bot}^4}{Q_s^2(x_2)}\exp\bigg(-\frac{\mathbf{k_\bot}^2}{Q_s^2(x_2)}\bigg)
\end{align}
which can be used to sample individual configurations of the color charge density as discussed below.

\subsection{Sampling of realistic color charge distributions}
When describing the color charge distributions of atomic nuclei, we follow the Monte Carlo Glauber Model and sample the position $x_{i}$ of the $i=1,\cdots,A$ individual nucleons according to a Wood-Saxon distribution. Each individual nucleon is assigned a three-dimensional thickness profile
\begin{align}\label{profile}
    T_{i}(x,y,z)=\frac{\gamma}{\sqrt{2\pi R_p^2}}e^{\frac{-(x-x_i)^2-(y-y_i)^2-(z-z_i)^2\gamma^2}{2R_p^2}}
\end{align}
such that the overall thickness of the nucleus is given by
\begin{eqnarray}
T(x,y,z)=\sum_{i=1}^{A} T_{i}(x,y,z)\;.
\end{eqnarray}
which according to Eq.~(\ref{eq:tmdresult}) is normalized such that $\int d^3x~T(x)= 2\pi R_{p}^2~A$, where $R_{p}=2 \rm{GeV}^{-1}$ is the proton radius. Since the spatial distribution $T$ typically varies on distance scales $\sim R_{p} \ll 1/Q_s$, the color charge distribution inside the nuclei is then sampled according to
\begin{align}\label{physicalcharge}
    \rho^a_{L/R}(t,x,y,z)=ga_xa_ya_z\sqrt{T(x,y,z)}\notag\\
    ~~\int d^2\mathbf{k_\bot} dk_z\sqrt{\tilde{\Gamma}(t,\mathbf{k_\bot},k_z)}
    \tilde{\zeta}^a(\mathbf{k_\bot},k_z) e^{i\vec{k}_\bot \vec{x}_\bot+k_zz}
\end{align}
where $\zeta(x_\bot,z)= 1/a^{3/2} \chi_{RNG}(x_\bot,z)$ are Gaussian random numbers and $\tilde{\zeta}(\mathbf{k_\bot},k_z)$ denotes their Fourier transform. Based on the initial conditions for the charge density profiles in Eq.~(\ref{physicalcharge}), we then proceed as described in Sec.~\ref{Sect2} to set up the initial conditions and simulate the dynamics of the collision. We note that due to the presence of longitudinal fluctuations, a finer discretization in the longitudinal ($z$) direction is required in this case, and unless stated otherwise, we will employ $Q_{0}a_{\perp}=0.33$ and $R_{p}/a_{z}=256$ in our numerical studies.

\subsection{Numerical results for realistic charges}
We now proceed to the study of the collision dynamics for realistic charge profiles, and consider head on $(b=0)$ Au-Au collisions for center of mass energies of 130 and 200 GeV. Since the basic features of the reaction dynamics remain essentially the same as for the simplistic charge profiles discussed in Sec.~\ref{Sect2}, we will focus our investigation on the violations of boost invariance and study the longitudinal fluctuations which emerge naturally within our framework.

We illustrate the full $3+1$ D structure in Fig.~\ref{3dRenderingsForEvolution}, which shows different phases of the collision for one particular event of a head-on Au-Au collision at $\sqrt{s_{NN}}=200$ GeV. In the left panel, we show the energy density $T^{00}_{WW}$  associated with the color fields of the incoming nuclei, which are well separated at $t=-0.37$ fm/c before the collision. Grey spheres overlaid to the energy density profile indicate the positions of nucleons, which dominate the longitudinal and transverse large scale structure of the energy density inside the nucleus. Small scale fluctuations of the color charge distribution results in additional inhomogeneities, clearly visible in the second nucleus. The central panel of Fig.~\ref{3dRenderingsForEvolution} shows the three dimensional profile of the transverse pressure $(T^{xx}+T^{yy})/2$ at the time of the collision $t=0$ fm/c i.e. when the incoming nuclei maximally overlap with each other. Since at this point, some of the individual nucleon-nucleon collisions have already taken plane, one starts to see formation of Glasma flux tubes of varying length, along with their fluctuations in the longitudinal and transverse direction. Subsequently, the Glasma flux tubes expand into longitudinal and transverse space, as can be seen from the right panel, showing the transverse pressure for a late time $t=0.6$ fm/c after the collision. Despite the fact that the individual flux tubes are stretched out along the longitudinal direction, as a consequence of fluctuation one observes clear non-uniformities of the flux tubes,  along with asymmetries in the forward and backward profiles. 
\begin{figure}%
    \centering
    \includegraphics[width=0.45\textwidth]{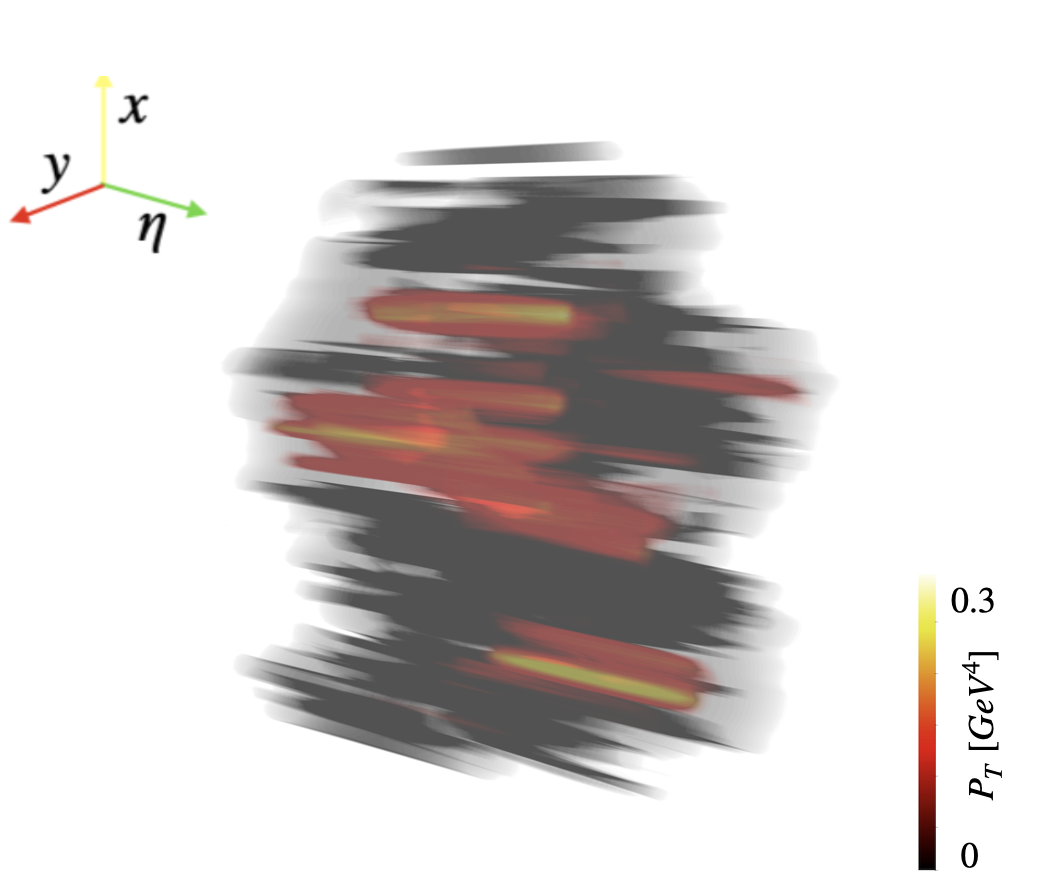}
    \caption{Three dimensional rendered view of transverse pressure at $\tau\simeq0.4$ fm/c in Milne coordinates for a single event of Au-Au collision at  $\sqrt{s_{NN}}=130$~GeV.}
    \label{3DRapidity}%
\end{figure}
\begin{figure}
    \centering
    \subfigure{{\includegraphics[width=0.38\textwidth]{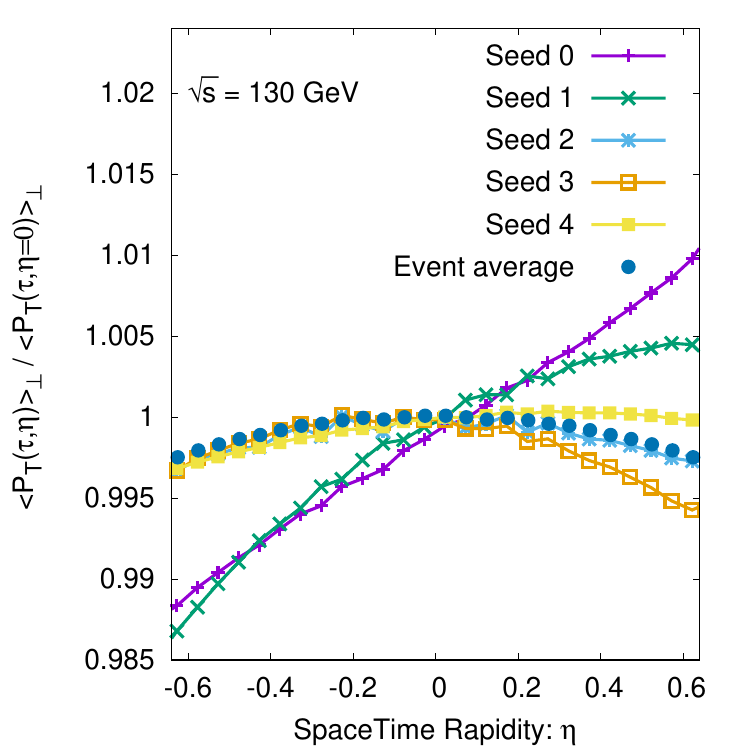}}}
    \subfigure{{\includegraphics[width=0.38\textwidth]{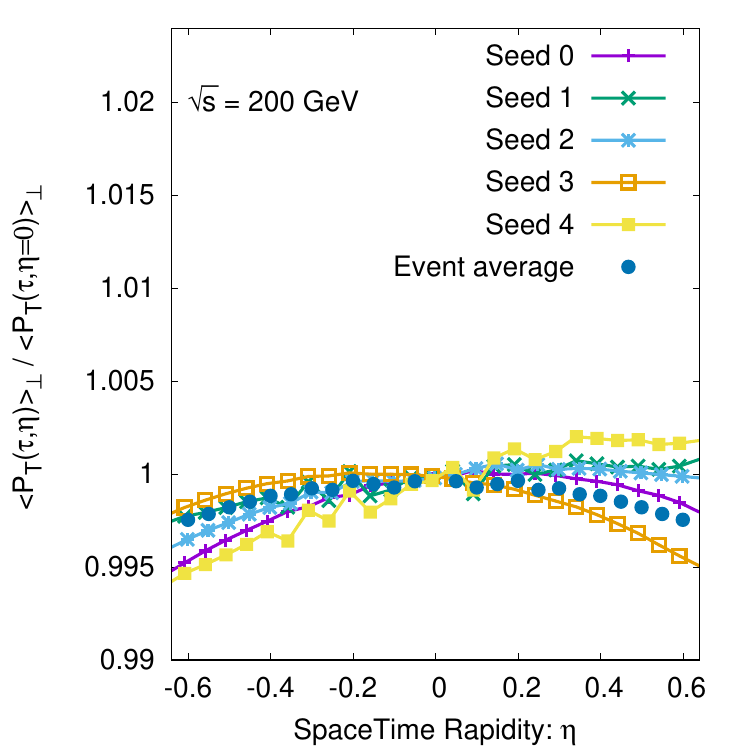}}}
    \caption{Integrated transverse pressure as a function of space-time rapidity for average and individual runs at center of mass energies of 130 and 200 GeV}%
    \label{BinnedRapidity}%
\end{figure}

Since at late times the rapid longitudinal expansion of the system stretches out the longitudinal profiles, some of the features of the Glasma are not clearly visible in the above figure and it is advantageous to visualize the corresponding structure in Milne coordinates. We illustrate this in Fig.~\ref{3DRapidity}, where we show the three dimensional profile for a head-on Au-Au collision at 130 GeV in $x,y$ and $\eta$ coordinates at $\tau\simeq0.4$~fm/c, which characterises the rapidity fluctuations. We note that due to limited availability of points near the light cone, we restrict ourselves to the central rapidity range $\eta \in [-0.8,0.8]$. Since at times $\tau \simeq 0.4$~fm/c the color charges of the colliding nuclei have escaped the central region, the transverse pressure profiles around mid rapidity are well described by approximately boost invariant flux tubes of varying transverse extent, and resemble the structures put forward in various models of the longitudinal structure of the initial state \cite{Broniowski:2016xnz,Pang:2015zrq}. By careful inspection of individual flux tubes in Fig.~\ref{3DRapidity} one also observes longitudinal fluctuations, albeit the amplitude of the longitudinal variations is significantly smaller compared to the variations in the transverse plane. 

Now in order to further analyze the longitudinal fluctuations, we consider the rapidity profiles of the transverse pressure $\langle P_{T}(\tau,\eta)\rangle_{\bot}$ averaged over the transverse plane. In Fig.~\ref{BinnedRapidity} we show the ratio of $\langle P_{T}(\tau,\eta)\rangle_{\bot}$ relative to its value $\langle P_{T}(\tau,\eta=0)\rangle_{\bot}$ at mid-rapidity at $\tau \simeq 0.75~$fm/c for head-on Au+Au collisions at two different energies $\sqrt{s_{NN}}=130,~200$~GeV in the top and bottom panels. Different curves in each panel correspond to the results for five individual events (labeled as Seed 0-4), along with the symmetrized average over all events. Generally, the fluctuations in the accessible rapidity window $-0.6<\eta<0.6$ are relatively small $\lesssim 1 \%$, and appear to decrease with increasing center of mass energy, as the longitudinal profile gets stretched out over a larger rapidity range. It is also interesting to observe that the dominant fluctuation in individual events around mid-rapidity appears to be a forward/backward asymmetry, and it would be interesting to extend this analysis to larger rapidities. However, this would require simulations on significantly finer and larger lattices, and is therefore beyond the scope of the present work.

\section{Conclusion \& Outlook}
We developed a framework to perform $3+1$D simulations of initial energy deposition in heavy-ion collision based on the effective theory of CGC, which takes the finite longitudinal extent of the colliding nuclei into account. Based on a simple model of the color charge distribution, we investigated the detailed dynamics, during and shortly after the collision. While in low energy collisions, where the longitudinal extent of the incoming nuclei $Q_sR_\gamma$ is non-negligible, significant violations of boost invariance can be observed, the results smoothly approach the boost invariant  limit \cite{Kovner:1995ts,Romatschke:2006nk} at high energies where the longitudinal thickness $Q_sR_\gamma \to 0$ becomes sufficiently small.

Subsequently, we developed a more elaborate model of the three dimensional color charge distribution in a large nucleus, where the large scale structure of the nucleus is determined by the longitudinal and transverse positions of nucleons, while small scale fluctuations in the longitudinal and transverse directions are determined by the $x$ and $\mathbf{k_\bot}$ dependence of transverse momentum dependent parton distributions. Based on this model, we obtained first results regarding the three dimensional structure and its fluctuations at two different center of mass energies, which show encouraging trends e.g. the longitudinal rapidity profiles and fluctuations appear to become stretched with increasing center of mass energy, which was not necessarily the case in a previous attempt to generalize the IP-Glasma initial state to $3+1$ dimensions \cite{Schenke:2016ksl}.

Due to the significant computational cost of performing $3+1$D classical Yang-Mills simulations of the space-time dynamics our numerical results have so far been limited to the central rapidity window for a few head-on collisions, and it would certainly be interesting to extend the analysis to larger rapidities and higher center of mass energies and perform a more systematic study of the various effects as a function of the centrality of the events. Evidently, to make contact with experimental observations, such phenomenological studies should be performed within the physical $SU(3)$ gauge group of QCD, where the formalism developed within this paper can be applied in exactly the same way, albeit further increasing the computational cost of the simulations. Beyond the improvement of numerical simulations (see also \cite{Ipp:2017lho}), it would also be important to develop further analytical insights into the $3+1$D space-time evolution of the Glasma, which could e.g. be obtained by analyzing the perturbative dilute limit along the lines of \cite{Altinoluk:2020wpf}.
\\

\textbf{ACKNOWLEDGMENTS:}
We thank A.~Dumitru, A.~Ipp, T.~Lappi, D.~Mueller, O.~Philipsen and B.~Schenke  for their valuable discussions. This work is supported by the DeutscheForschungsgemeinschaft (DFG, German Research Foundation)
through the CRC-TR 211 ’Strong-interaction matter under extreme conditions’– project
number 315477589 – TRR 211. The computation done in this work was performed at Paderborn Center for Parallel Computing (PC2
) and National Energy Research Scientific Computing Center (NERSC), a DOE Office of Science User Facility supported by the Office of
Science of the U.S. Department of Energy under Contract No. DE-AC02-05CH11231

\begin{appendix}
\begin{figure*}
    \centering
    \subfigure{{\includegraphics[width=0.24\textwidth]{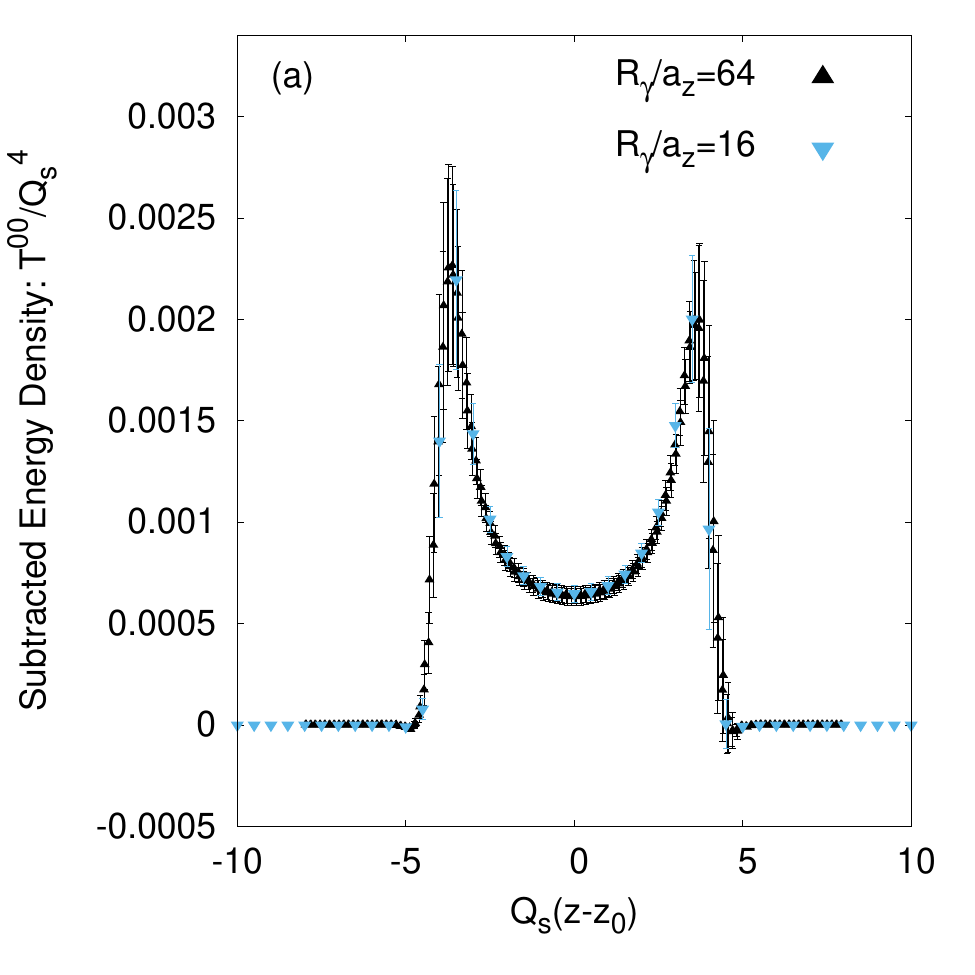}}}%
    \subfigure{{\includegraphics[width=0.24\textwidth]{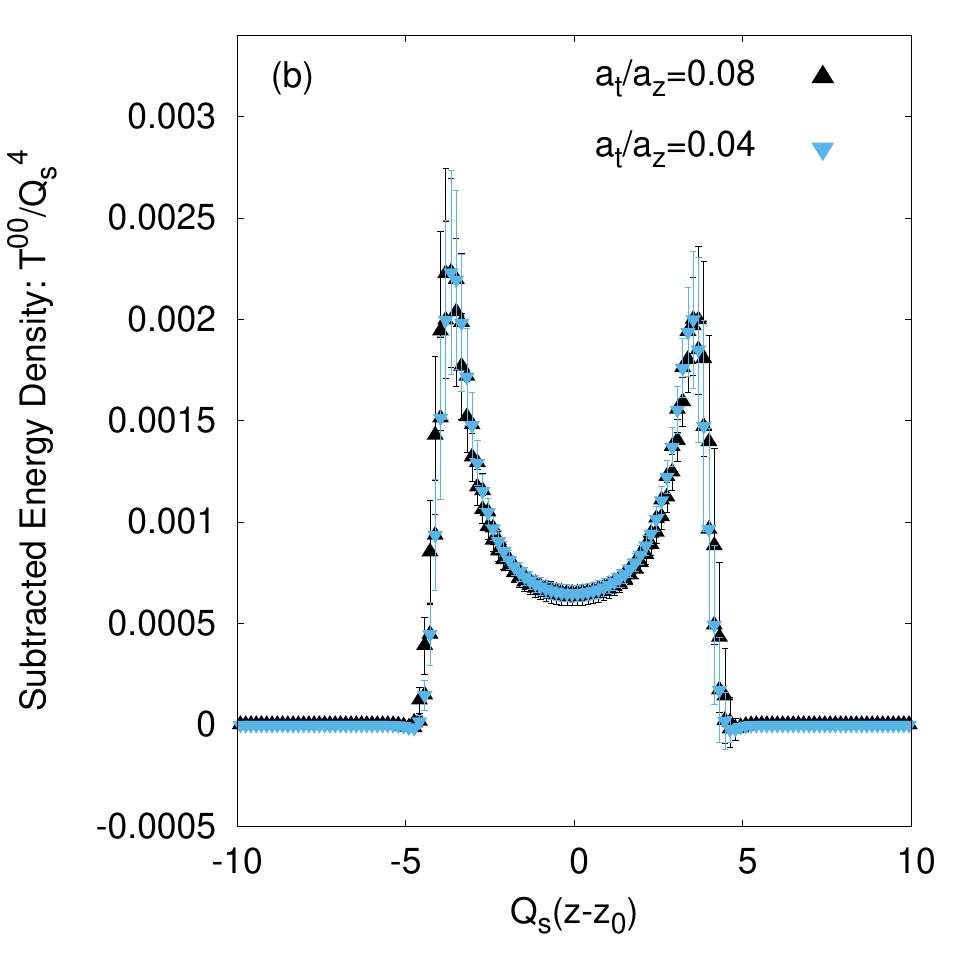}}}
    \subfigure{{\includegraphics[width=0.24\textwidth]{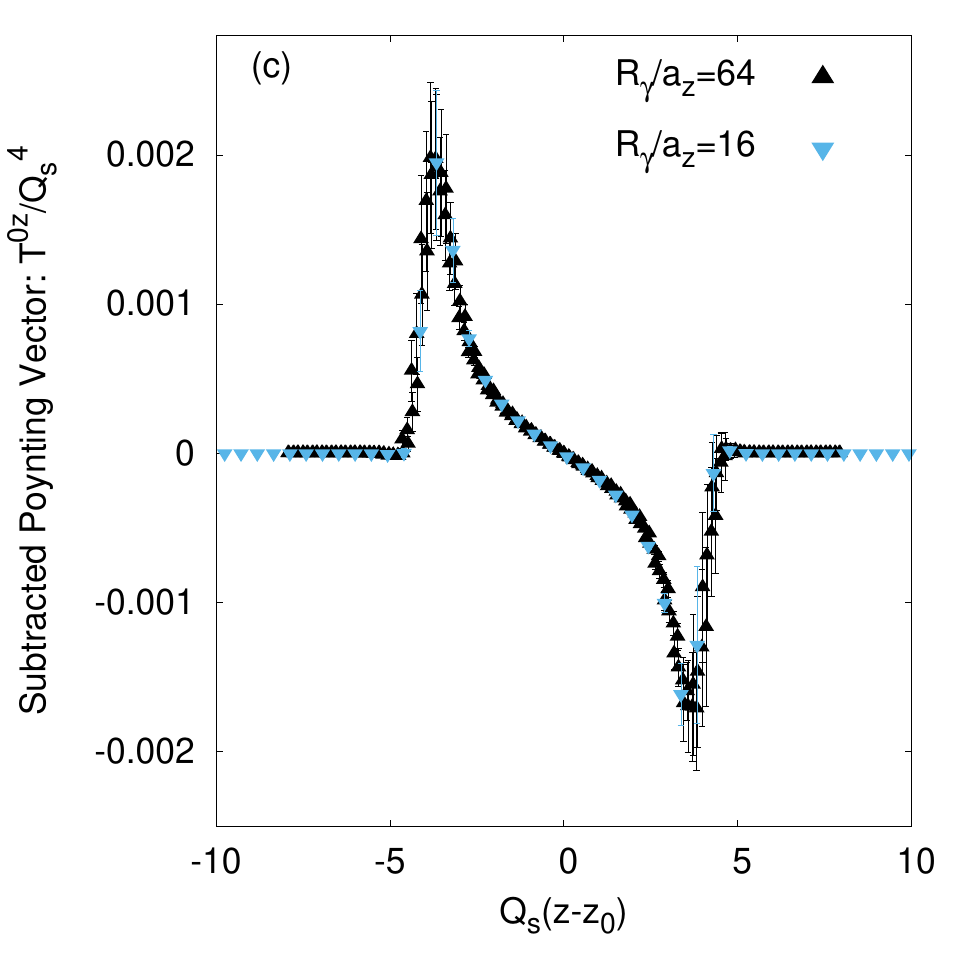} }}%
    \subfigure{{\includegraphics[width=0.24\textwidth]{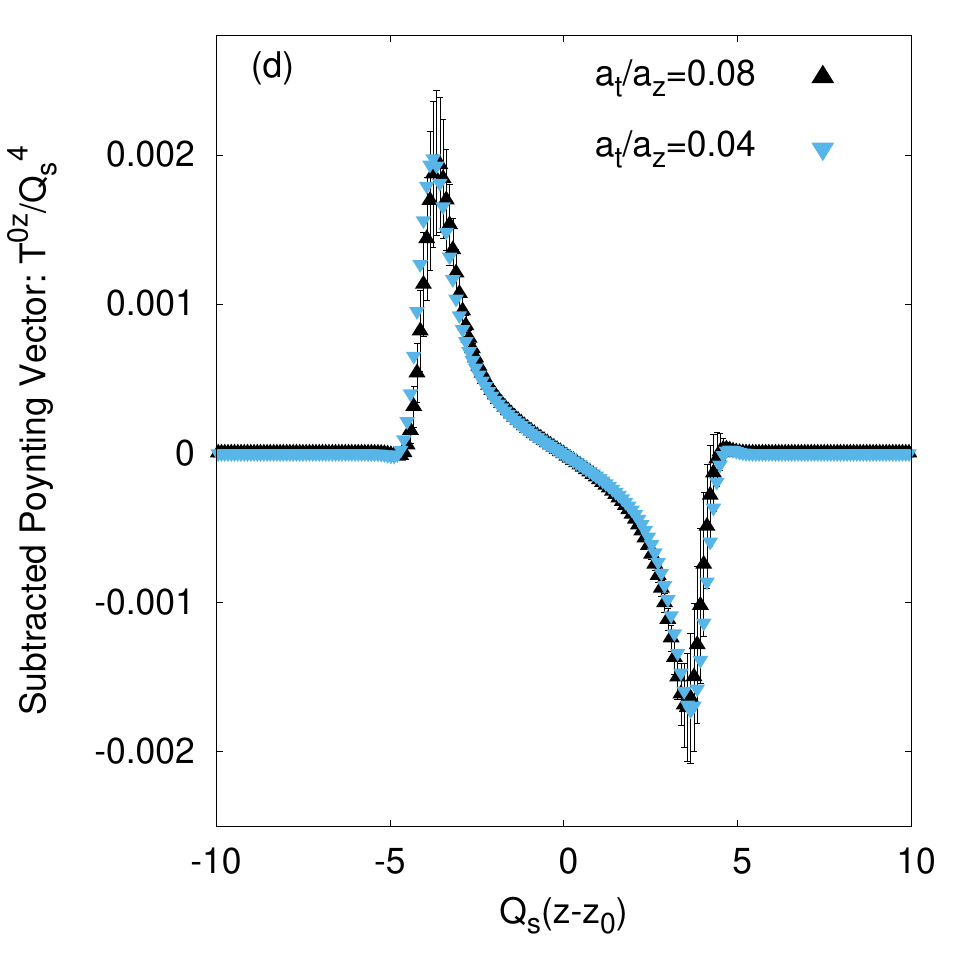} }}
    \caption{Discretization effects on the various components of the energy momentum tensor for $Q_sR_\gamma=0.5$ at $Q_st \simeq 4.$ Effect of finer longitudinal lattice spacing $a_z \to 0 $ is shown in panel (a) and (c) whereas that of finer time discretization $a_t/a_z \to 0 $ is shown in panel (b) and (d).}%
    \label{SpuriousFields}%
\end{figure*}
\section{Discretization effect and approach to the continuum limit}
\label{Appendix:Continnumlimit}
Below we provide additional results for simulations where we vary the various discretization parameters, to illustrate the results presented in the main part of this work do not suffer from significant discretization artifacts.

We present a compact summary of the results in Fig.~\ref{SpuriousFields}, where for a fixed value of $Q_st \simeq 4$, subtracted $T^{00}$ and $T^{0z}$ components of the energy momentum tensor are shown for the thick nuclei $(Q_sR_\gamma=0.5)$. In the first two panels: $T^{00}$ obtained with the lattice discretization used throughout the manuscript $(R_\gamma/a_z=16;   a_t/a_z=0.08)$ is compared against the values obtained with finer longitudinal lattice spacing $(R_\gamma/a_z=64)$ and finer time step $(a_t/a_z=0.04).$ Lattice dimension is taken to be $128^2\times 1024$ and $128^2\times 2048$ for $R_\gamma/a_z=16$ and $R_\gamma/a_z=64$ respectively. We observe that approaching the continuum limit doesn't shrink the spurious fields in the proximity of light cone. 
In the other two panels, similar result is shown for  $T^{0z}$, which again shows that the effect of these contributions do not change with finer lattice spacing.
\end{appendix}
\bibliographystyle{unsrtnat}
\bibliography{references}
\end{document}